\documentclass[twocolumn]{aastex631}
\submitjournal{Accepted by NST}

\shorttitle{The ws-process in massive stars}
\shortauthors{Xin et al.}

\graphicspath{{./}{figures/}}
\usepackage{courier}
\usepackage{multirow}
\usepackage{threeparttable}
\usepackage{booktabs}
\usepackage{color}
\usepackage[section]{placeins}
\usepackage{graphicx}
\usepackage{subfigure}
\usepackage{parskip}
\usepackage{longtable}
\usepackage{supertabular}
\usepackage{bm}

\begin{document}

\title{The impact of new ($\alpha$, n) reaction rates on the weak s-process
in metal-poor massive stars}

\correspondingauthor{Wenyu Xin, Xianfei Zhang}
\email{xinwenyu16@mails.ucas.ac.cn, 11112017095@bnu.edu.cn}

\author[0000-0003-3646-9356]{Wenyu Xin}

\affiliation{Institute for Frontiers in Astronomy and Astrophysics, Beijing Normal University, Beijing 102206, China}
\affiliation{School of Physics and Astronomy, Beijing Normal University, Beijing 100875, People's Republic of China}

\author[0000-0002-3311-5387]{Chun-Ming Yip}
\affiliation{GSI Helmholtzzentrum f\"ur Schwerionenforschung, Planckstra{\ss}e 1, D-64291 Darmstadt, Germany}
\affiliation{Department of Physics and Institute of Theoretical Physics, The Chinese University of Hong Kong, Shatin, N.T., Hong Kong S.A.R., People’s Republic of China}

\author[0000-0001-9553-0685]{Ken'ichi Nomoto}
\affiliation{Kavli Institute for the Physics and Mathematics of the Universe (WPI),
The University of Tokyo Institutes for Advanced Study, \\
The University of Tokyo, Kashiwa, Chiba 277-8583, Japan}

\author[0000-0002-3672-2166]{Xianfei Zhang}
\affiliation{Department of Astronomy, Beijing Normal University, Beijing 100875, China}
\affiliation{Institute for Frontiers in Astronomy and Astrophysics,
Beijing Normal University, Beijing 102206, China}

\author[0000-0002-7642-7583]{Shaolan Bi}
\affiliation{Department of Astronomy, Beijing Normal University, Beijing 100875, China}
\affiliation{Institute for Frontiers in Astronomy and Astrophysics,
Beijing Normal University, Beijing 102206, China}

%\author[0000-0002-8980-945X]{Gang Zhao}
%\affiliation{CAS Key Laboratory of Optical Astronomy, National Astronomical Observatories,
%Chinese Academy of Sciences, Beijing 100101, China}
%\affiliation{School of Astronomy and Space Science, University of Chinese Academy of
%Sciences, Beijing 100101, China}



\begin{abstract}

Massive stars are significant sites for the weak s-process (ws-process). 
$^{22}$Ne and $^{16}$O are, respectively, the main neutron source and poison for the ws-process.
In the metal-poor stars, the abundance of $^{22}$Ne is limited by the
metallicity, so that the contribution of $^{22}$Ne($\alpha$, n)$^{25}$Mg
reaction on the s-process is weaker.
Conversely, the $^{17}$O($\alpha$, n)$^{20}$Ne reaction becomes more prominent
in these stars due to the most abundant $^{16}$O in all metallicities.  
In this work, we calculate the evolution of four metal-poor
models ($Z=10^{-3}$) for the Zero-Age Main-Sequence (ZAMS) masses of
$M ({\rm ZAMS})=$ 15, 20, 25, and 30 M$_{\odot}$ to investigate the
effect of reaction rates on the ws-process.
We adopt the new $^{17}$O($\alpha$, n)$^{20}$Ne and $^{17}$O($\alpha, \gamma$)$^{21}$Ne
reaction rates suggested by Best et al. (2013) and $^{22}$Ne($\alpha$, n)$^{25}$Mg 
and $^{22}$Ne($\alpha, \gamma$)$^{26}$Mg from Wiescher et al. (2023). 
The yields of the s-process isotope with updated reaction rates
are compared with the results using default reaction rates from JINA REACLIB.
We find that the new $^{17}$O+$\alpha$ reaction rates increase the ws-process
mainly in all the stages, while the new $^{22}$Ne+$\alpha$ reaction rates only
increase the ws-process in C and Ne burning stages. Updating these new reaction
rates would increase the production of ws-process isotopes by tens of times.
We also note that for more massive stars, the enhancement by new $^{17}$O+$\alpha$
reaction rates become more significant.

\end{abstract}



\section{Introduction} \label{sec:intro}

Massive stars play crucial roles in galactic chemical evolution by
synthesizing elements up to the iron group through
charged-particle reactions during thermonuclear burning.
The slow neutron capture process (s-process),
produces heavy elements in stars by allowing atomic nuclei to capture
neutrons at a rate slow enough to allow unstable isotopes to undergo
beta decay before capturing additional neutrons.

In massive stars with Zero-Age Main-Sequence (ZAMS) masses greater than
$\sim$ 12 $M_{\odot}$, the weak s-process (ws-process) is a key
mechanism for producing neutron-rich isotopes, particularly those
in the atomic mass range of $A = $ 60 to 90 \citep{1989RPPh...52..945K}.
Early studies associated the ws-process primarily with core helium (He) burning
\citep{1974ApJ...190...95C, 1985ApJ...295..589A, 1989A&A...210..187L, 
1990A&A...234..211P, 1992A&A...258..357B, 2000ApJ...533..998T}.
Later research identified significant production during shell carbon (C) burning,
which is characterized by higher temperatures and neutron densities
\citep{1991ApJ...371..665R, 1992ApJ...387..263R, 1993ApJ...419..207R, 
2007ApJ...655.1058T}.
More recent models include explosive nucleosynthesis during
core-collapse supernovae (CCSNe),
although these events have minimal impact on ws-process yields
\citep{2001ApJ...549.1085H, 2002ApJ...576..323R,
2003ApJ...592..404L, 2009ApJ...702.1068T, Limongi_2018}.
\citet{2003ApJ...592..404L} and \citet{2009ApJ...702.1068T} showed that
the yields of the ws-process are not strongly modified by the supernova explosion.

In contrast to the main s-process in asymptotic giant branch (AGB) stars,
which relies on the $^{13}$C($\alpha$, n)$^{16}$O reaction,
the ws-process in massive stars is driven by the $^{22}$Ne($\alpha$, n)$^{25}$Mg
\citep{1968ApJ...154..225P, 1974ApJ...190...95C, 1990A&A...234..211P, 1991ApJ...367..228R}.
$^{22}$Ne in core He burning is produced via a sequence of reactions,
$^{14}$N($\alpha, \gamma$)$^{18}$F($\beta^+ \nu$)$^{18}$O($\alpha, \gamma$)$^{22}$Ne.
The ws-process is activated by the $^{22}$Ne($\alpha$, n)$^{25}$Mg reaction once the
temperature exceeds 2.0$\times10^{8}$ K.
During shell C burning, this reaction is re-activated by $\alpha$ produced
by the $^{12}$C($^{12}$C, $\alpha$)$^{20}$Ne channel \citep{1969ApJ...157..339A}.

$^{22}$Ne is primarily synthesized by $\alpha$-capture involving $^{14}$N,
whose amount depends on the initial metallicity of stars.
Consequently, one would expect the yields of the ws-process elements to be low
in metal-poor stars \citep{1995ApJS..101..181W, 1992A&A...258..357B}.
However, recent observations found that the ws-process elements
in metal-poor stars are not as low as previously predicted
\citet{2005ApJ...632..611A, 2006ApJ...639..897A, 2011Natur.472..454C}.
To account for this discrepancy, theoretical models have proposed that
fast-rotating massive stars may enhance the production of the ws-process elements.
In these models, rotation can promote the mixing of $^{14}$N from the H-rich
envelope into the convective core of He burning and increase their production
\citep{2011Natur.472..454C, 2012A&A...538L...2F, Limongi_2018}. 

Moreover, uncertainties in $^{17}$O+$\alpha$ reaction rates significantly affect
the yields of the ws-process, particularly in metal-poor stars,
where $^{16}$O acts as a major neutron poison through
$^{16}$O(n, $\gamma$)$^{17}$O reaction \citep{2010ApJ...710.1557P}.
Subsequent competing reactions 
$^{17}$O($\alpha$, n)$^{20}$Ne and $^{17}$O($\alpha$, $\gamma$)$^{21}$Ne determine 
whether neutrons are released or captured.
Although recent studies have explored these effects in rotating stars
\citep{2012A&A...538L...2F, 2017MNRAS.469.1752N, 2018A&A...618A.133C},
few have investigated the combined impact of $^{17}$O+$\alpha$ and $^{22}$O+$\alpha$
reactions in non-rotating metal-poor stars.
Since $^{16}$O is extremely abundant across all metallicities,
the neutrons released by the $^{22}$Ne($\alpha$, n)$^{25}$Mg reaction
in metal-poor stars are possibly captured by $^{16}$O,
instead of attending the ws-process,
and followed by a large amount of production of $^{17}$O.
Therefore, the $^{17}$O($\alpha$, n)$^{20}$Ne reaction could play a
more important role.

In this study, we investigate the ws-process in non-rotating massive stars,
specifically comparing these new reaction rates suggested in recent references
with those in JINA REACLIB \citep{2010ApJS..189..240C}.
We evaluate the implications of these new reaction rates on the ws-process,
emphasize how variations in these rates influence nucleosynthesis.
In Section \ref{sec:model}, we present the parameters of our stellar models and compare
the results using the reaction rates from the new references with those from JINA REACLIB.
In Section \ref{sec:evo_exp}, we use a model with $M(\rm ZAMS)$ = 25 M$_\odot$ as an
example to illustrate the evolution of metal-poor stars.
We further compare the effects of the $^{17}$O+$\alpha$ and $^{22}$Ne+$\alpha$ reactions
on nucleosynthesis in Section \ref{sec:nuc}.
We conclude the study in Section \ref{sec:conclusion}.

\section{Models and Input Physics}  \label{sec:model}

We employ the Modules for Experiments in Stellar Astrophysics
(MESA, version 12778; \citet{2011ApJS..192....3P, 2013ApJS..208....4P,
2015ApJS..220...15P, 2018ApJS..234...34P, 2019ApJS..243...10P, 2023ApJS..265...15J})
to follow various nuclear burnings and the structural evolution
in stars from ZAMS until the Fe core collapse,
when the infall velocity of the Fe core reaches 10$^3$ km s$^{-1}$.
We only focus on nucleosynthesis before the explosion,
as the final explosion makes only a slight modification of the
ws-process abundance \citep{2009ApJ...702.1068T}.

We calculate the evolution of four metal-poor stellar models of 
$M ({\rm ZAMS})=$ 15, 20, 25, and 30 M$_{\odot}$ with MESA
\footnote{All these models are available on Zenodo at doi:\href{https://doi.org/10.5281/zenodo.17021896}{10.5281/zenodo.17021896}}.
The trajectories of these models are utilized in the WinNet code
\citep{2023ApJS..268...66R}
to investigate the effects of reaction rates on the ws-process.
For the $^{17}$O+$\alpha$ reactions, we incorporate both competing reactions, 
$^{17}$O($\alpha$, n)$^{20}$Ne and $^{17}$O($\alpha, \gamma$)$^{21}$Ne,
as reported by \citet{2013PhRvC..87d5805B}.
The reaction rates for $^{22}$Ne+$\alpha$, 
including both the  ($\alpha$, n) and ($\alpha$, $\gamma$) reactions,
are updated according to \citet{2023EPJA...59...11W}.
To assess the impact of these reactions,
we compare four reaction recipes for each model,
as listed in Table~\ref{tab:recipes}. The differences among these reaction
rates are discussed in Section \ref{sec:rates}.
Most physical parameters follow \citet{2023ChPhC..47c4107X, 2025arXiv250211012X}
with some changes clarified in Section \ref{sec:input}.
Section \ref{sec:WinNet} outlines the setup within the WinNet code.

\begin{table}[hbtp]
\centering
\caption{The reaction recipes used in this work.}
\label{tab:recipes}
\begin{threeparttable}
\begin{tabular}{ccc}
\toprule
  Case & $^{17}$O+$\alpha$     & $^{22}$Ne+$\alpha$    \\
\midrule
1 & REACLIB\tnote{*}      & REACLIB  \\
2 & REACLIB             & \citet{2023EPJA...59...11W} \\
3 & \citet{2013PhRvC..87d5805B} & REACLIB \\
4 & \citet{2013PhRvC..87d5805B} & \citet{2023EPJA...59...11W} \\
\bottomrule
\end{tabular}
\begin{tablenotes}
\footnotesize
\item[*] JINA REACLIB \citet{2010ApJS..189..240C}.
\end{tablenotes}
\end{threeparttable}
\end{table}

\begin{figure}[htbp]
\centering
\begin{minipage}[c]{0.48\textwidth}
\includegraphics [width=80mm]{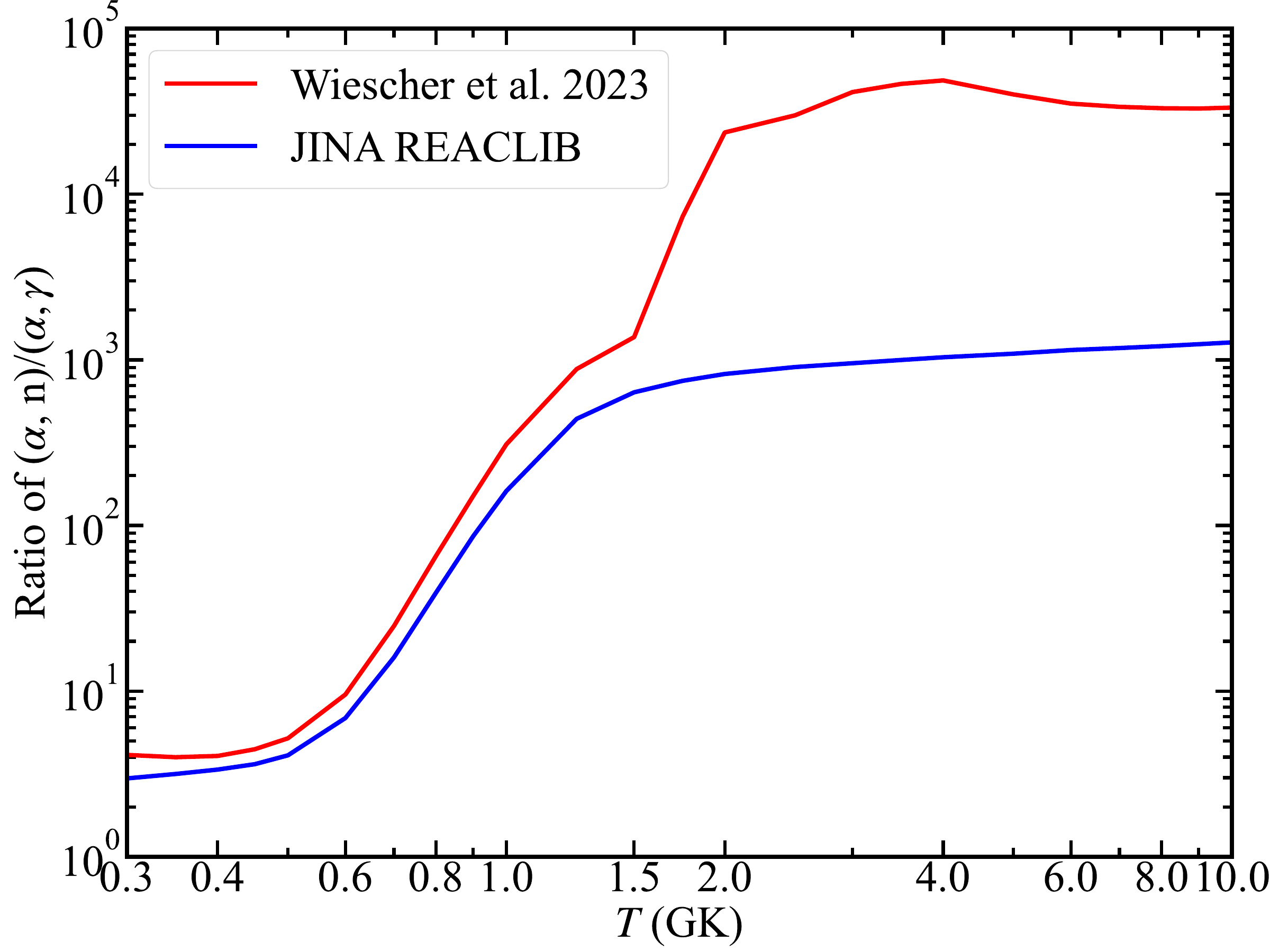}
%\centerline{(a) }
\end{minipage}%
\begin{minipage}[c]{0.48\textwidth}
\includegraphics [width=80mm]{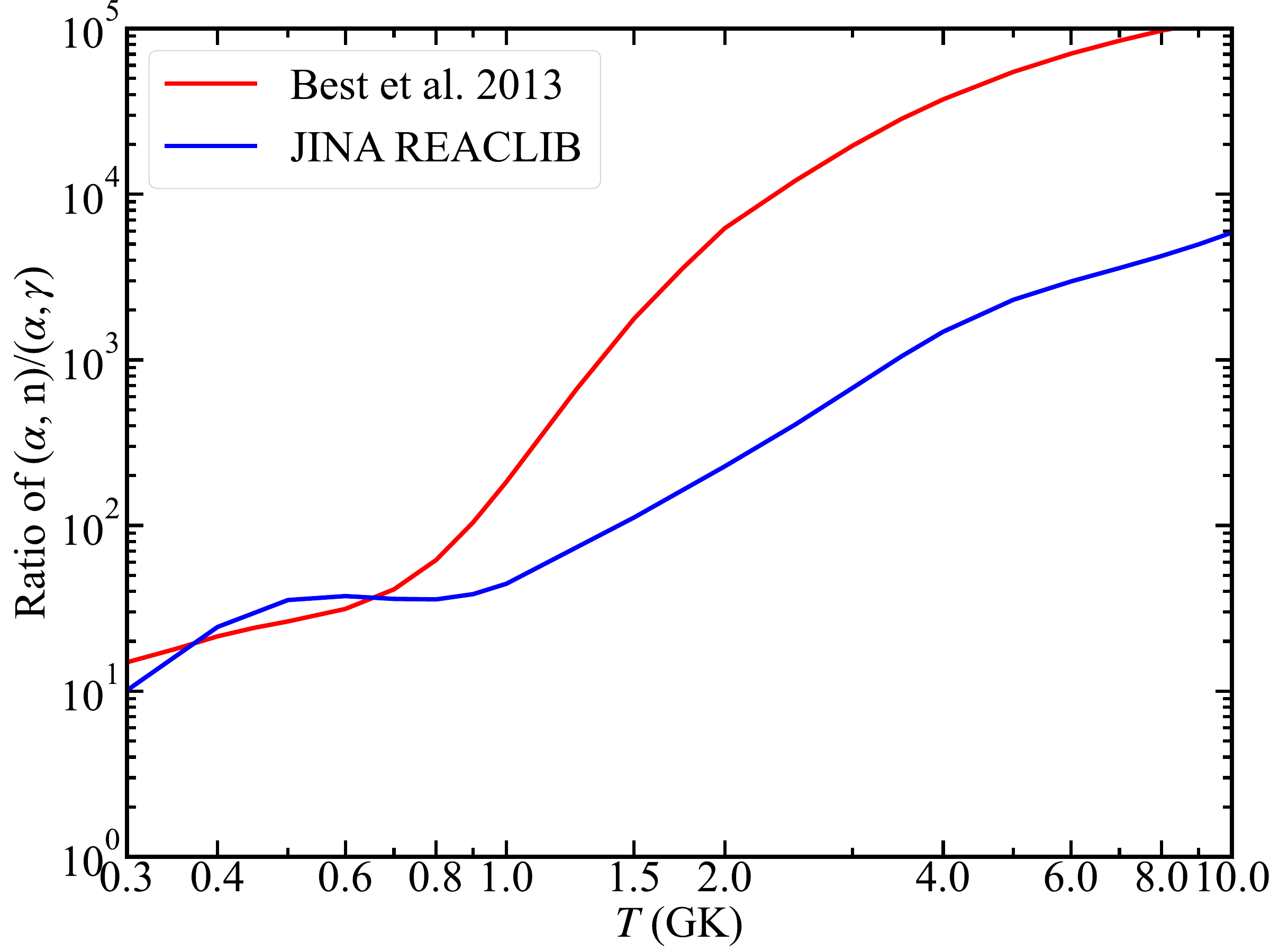}
%\centerline{(a) }
\end{minipage}%
\caption{The ($\alpha$, n)/($\alpha, \gamma$) ratio as a function of temperature
for $^{22}$Ne+$\alpha$ (top) and $^{17}$O+$\alpha$ (bottom) reactions.
\label{fig:rates_ratio}}
\end{figure}

\subsection{Reactions for Weak s-process}  \label{sec:rates}

The $^{22}$Ne($\alpha$, n)$^{25}$Mg reaction is active at $T \sim$
0.2 GK and 1.0 GK in He and C burning shells, respectively.
This reaction competes with $^{22}$Ne($\alpha$, $\gamma$)$^{26}$Mg, 
which consumes $^{22}$Ne without releasing neutrons.
In these shells, $^{16}$O is the most abundant isotope and acts as the
main neutron poison through $^{16}$O(n, $\gamma$)$^{17}$O.
Fortunately, the neutrons absorbed by $^{16}$O can be released again via
$^{17}$O($\alpha$, n)$^{20}$Ne. 
Therefore, the availability of neutrons for the ws-process is determined by the 
($\alpha$, n)/($\alpha$, $\gamma$) ratio for both 
$^{22}$Ne+$\alpha$ and $^{17}$O+$\alpha$ reactions.

In Figure \ref{fig:rates_ratio}, we show the ($\alpha$, n)/($\alpha$, $\gamma$)
ratios for the $^{22}$Ne+$\alpha$ (top panel) and $^{17}$O+$\alpha$ (bottom panel)
reactions as a function of the temperature. In the top panel,
the ($\alpha$, n)/($\alpha$, $\gamma$) ratio for $^{22}$Ne+$\alpha$,
as recommended by \citet{2023EPJA...59...11W}, is observed to be 1.2 to 2.0 times
higher than the values provided by REACLIB below 1.5 GK,
a range typically associated with He and C shell burning.
Notably, this enhancement increases dramatically,
reaching several tens of times above 1.5 GK.

In the bottom panel, the ($\alpha$, n)/($\alpha$, $\gamma$) ratio for
the $^{17}$O+$\alpha$ reaction suggested by \citet{2013PhRvC..87d5805B}
is similar to REACLIB below 0.7 GK, where only He burns,
but, this ratio rapidly increases to several tens of times in
the C, Ne, and O layers. With these updated reaction rates,
we anticipate an increase in the neutron release
from $^{22}$Ne while reducing neutron consumption by $^{16}$O. 
Consequently, the yields of the ws-process isotopes are significantly enhanced.

\subsection{Input Physics in MESA}  \label{sec:input}

\begin{table}[htb]
\centering
\caption{Nuclides included in the nuclear reaction network of \texttt{mesa\_161.net}.}
\label{tab:network}
\begin{tabular}{cccccc}
\toprule
Element & $A_{\rm min}$ & $A_{\rm max}$ & Element & $A_{\rm min}$ & $A_{\rm max}$ \\
\midrule
n       & 1    & 1    & S       & 31   & 35   \\
H       & 1    & 2    & Cl      & 35   & 38   \\
He      & 3    & 4    & Ar      & 35   & 40   \\
Li      & 7    & 7    & K       & 39   & 44   \\
Be\tnote{1}      & 7    & 10   & Ca      & 39   & 46   \\
B       & 8    & 8    & Sc      & 43   & 48   \\
C       & 12   & 13   & Ti      & 43   & 51   \\
N       & 13   & 15   & V       & 47   & 53   \\
O       & 14   & 18   & Cr      & 47   & 57   \\
F       & 17   & 19   & Mn      & 51   & 57   \\
Ne      & 18   & 22   & Fe      & 51   & 61   \\
Na      & 21   & 24   & Co      & 55   & 63   \\
Mg      & 23   & 26   & Ni      & 55   & 64   \\
Al      & 25   & 28   & Cu      & 59   & 64   \\
Si      & 27   & 31   & Zn      & 60   & 64   \\
P       & 30   & 33   &         &      &      \\
\bottomrule
\end{tabular}
\begin{tablenotes}
\footnotesize
\item[1] $^{8}$Be is not included.
\end{tablenotes}
\end{table}

To achieve the convergence of the model structures within approximately 10\%,
a nuclear network comprising at least 127 isotopes should be included
\citep{2016ApJS..227...22F}. In this work,
we utilize a more extensive nuclear network (\texttt{mesa\_161.net}) that
incorporates additional neutron-rich isotopes.
Table \ref{tab:network} lists all the isotopes in \texttt{mesa\_161.net}.
We adopt a metallicity of $Z=0.1\, Z_\odot$ and assume solar
metallicity ratios based on the work of \citet{1989GeCoA..53..197A}.

We have enhanced both the temporal and spatial resolutions to ensure the 
numerical convergence. The mass resolution is critical for accurately capturing
changes in the stellar structure \citep{2016ApJS..227...22F, 2022ApJ...937..112F}.
The parameter \texttt{max\_dq} controls the maximum fractional mass of a cell
in the model, and we set \texttt{max\_dq=5d-4}, which results in over 3,500 cells
in the model. We adopt a minimum diffusion coefficient of 
$D_{\rm min}=$ 10$^{-2}$ cm$^2$ s$^{-1}$ to ensure that the global mixing timescale
($\tau=L^2/D_{\rm min}$) is significantly longer than the lifetimes of
the stellar models. This allows us to neglect the effects of global mixing and
to smooth local composition gradients \citep{2022ApJ...937..112F}.

After C burning, the core structure becomes more complex
because of multi-shell burning, with the central entropy being significantly
influenced by shell burning (see \citet{2025arXiv250211012X}). 
To achieve finer granularity during the evolution, we impose limits on the changes in
the logarithm of the central density and temperature. 
Specifically, we set $\delta_{\rm log \, \rho_c}<10^{-3}$
and $\delta_{{\rm log} \, T_c}<2.5 \times 10^{-3}$.
Additionally, we restrict the change in the mass fraction of isotopes with
\texttt{dX\_nuc\_drop\_limit}$=3\times 10^{-2}$ and tightening this limit to
\texttt{dX\_nuc\_drop\_limit\_at\_high\_T}$=10^{-2}$ when log $T_{\rm c}>$ 9.45.

\subsection{Post-processing Calculation with WinNet} \label{sec:WinNet}

The detailed nucleosynthesis in the stellar models is computed in post-processing
using an extensive nuclear reaction network code WinNet \citep{2023ApJS..268...66R}.
The network consists of about 2000 isotopes from neutron and proton to thorium ($Z$ = 90).
The reaction rates of $(n,\gamma)$, $(n,p)$, $(p,\gamma)$, $(\alpha,n)$,
$(\alpha,p)$, $(\alpha,\gamma)$, and their inverse reactions from the JINA REACLIB
database \citep{2010ApJS..189..240C} are included.
Theoretical weak rates from \citet{2001ADNDT..79....1L},
electron chemical potentials from \citet{1999ApJS..125..277T},
and screening corrections from \citet{2014PhRvC..89a5802K} are used.
%https://cococubed.com/code_pages/chemical_potential.shtml

For each stellar model, we map the initial composition
and time evolutions of temperature and density from the MESA
simulation on trajectories.
The nucleosynthesis calculation of these trajectories is performed until
the onset of iron core-collapse.
The region inside the steepest-density jump 
is expected to eventually collapse into a neutron star  
and does not contribute to the yield of the ws-process nucleosynthesis.
The steepest density jump occurs at the most active burning shell
and has been defined in \citet{2025arXiv250211012X}.
We will describe briefly the MESA result in Section \ref{sec:evo_exp}.

\section{Evolution of Massive Stars} \label{sec:evo_exp}

\begin{figure}[htbp]
\centering
\begin{minipage}[c]{0.48\textwidth}
\includegraphics [width=80mm]{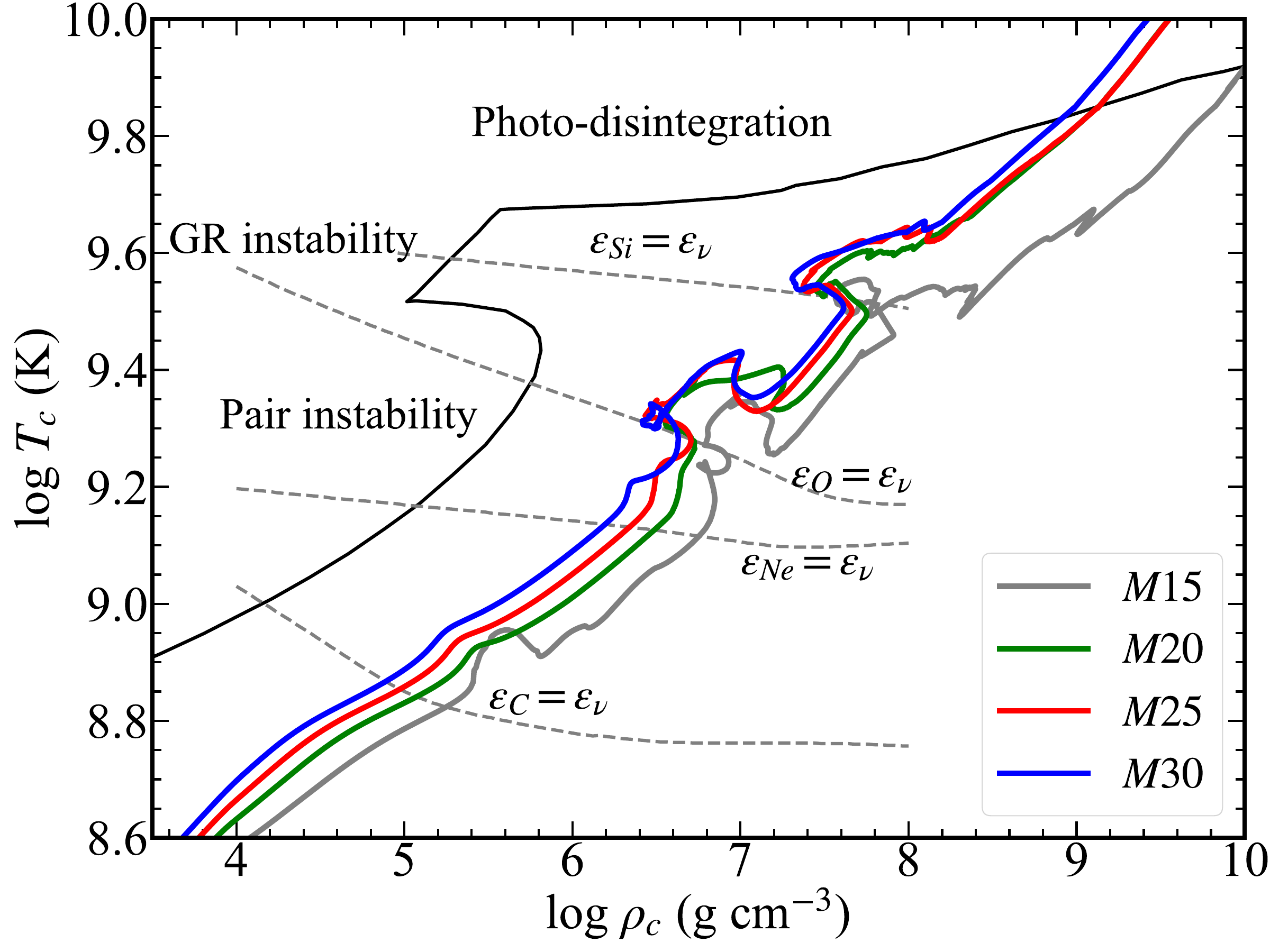}
%\centerline{(a) }
\end{minipage}%
\caption{The central temperature against the central density for the evolution of
stars with $M ({\rm ZAMS}) =$ 15, 20, 25 and 30 M$_{\odot}$.
The grey dashed lines show the ignition lines of C burning, Ne burning, O burning
and Si burning, where the energy generation rate by nuclear burning equals the energy
loss rate by neutrino emissions.
In the region on the left of the black line, stars are dynamically unstable
due to the electron-positron pair creation (indicated as ``pair instability'') (\citealt{2009ApJ...706.1184O}), general relativistic effects (``GR instability'') 
(see, e.g., \citealt{1966PASJ...18..384O}),
and the photo-disintegration of matter in nuclear
statistical equilibrium (NSE) at $Y_{\rm e}=$ 0.5 (``photo-disintegration'') (\citealt{2009ApJ...706.1184O}).
\label{fig:trho}}
\end{figure}

\begin{figure}[h]
\centering
\begin{minipage}[c]{0.45\textwidth}
\includegraphics [width=80mm]{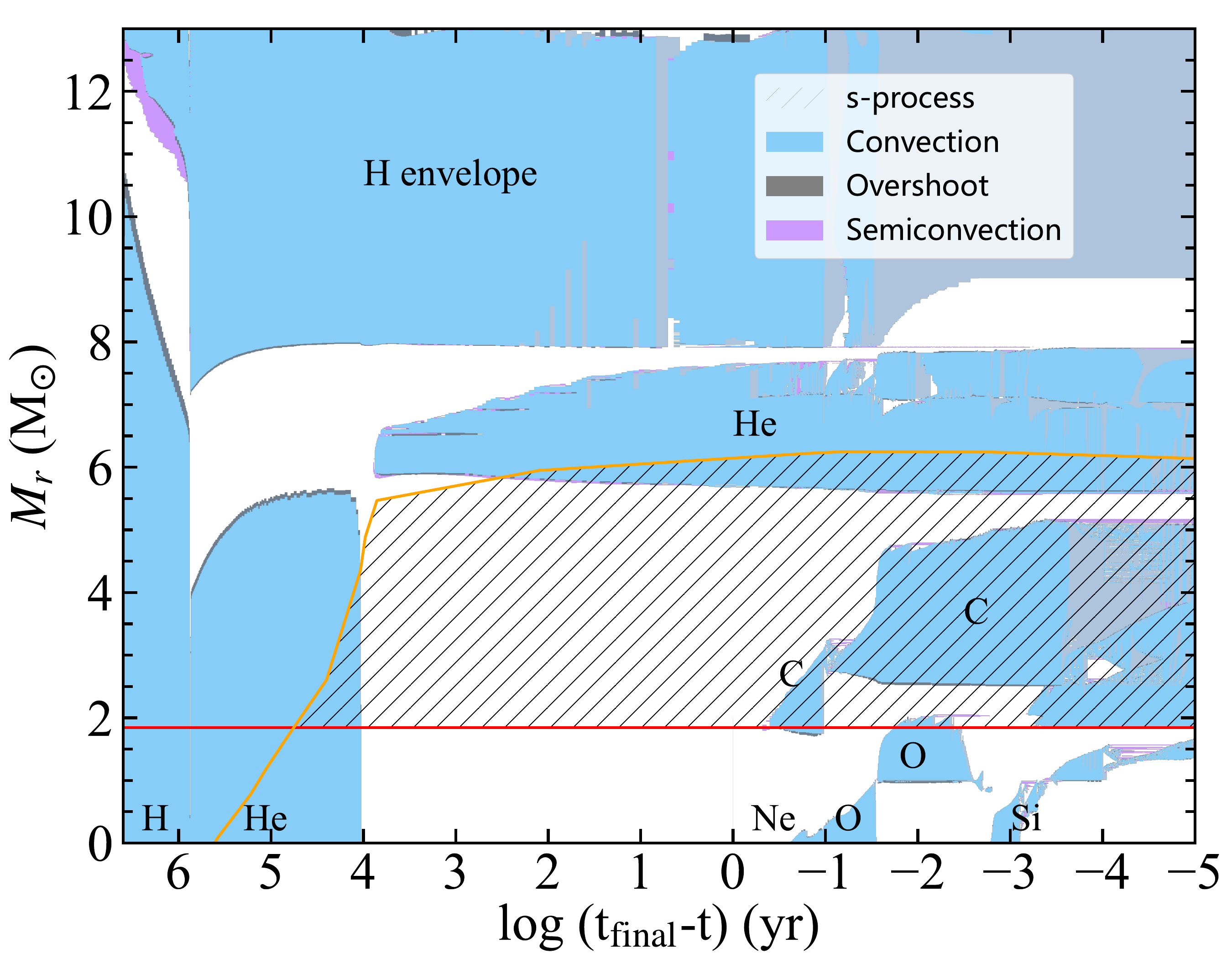}
%\centerline{(a) }
\end{minipage}%
\caption{The Kippenhahn diagram of the star with $M {\rm (ZAMS)}$ = 25 M$_{\odot}$
The inner part of $M_r = 0 - 14$ M$_{\odot}$ is shown.
The blue, grey, and pink represent the convection, overshoot, and semiconvection
regions, respectively. The orange line is the isotherm line of 0.2 GK, and the
red line shows the location of $M_r=1.84$ M$_\odot$. Between these two lines,
the hatched region indicates where the ws-process is taken into consideration.
\label{fig:25M_evo}}
\end{figure}

After core He burning, the mass fraction of $^{12}$C in the center
is smaller for larger $M ({\rm ZAMS})$.
Only the star with $M ({\rm ZAMS}) =$ 15 M$_{\odot}$ can ignite
convective C burning in the center, as it has sufficient 
fuel with $X$($^{12}$C) $\sim$ 0.2.
In contrast, other models with larger $M ({\rm ZAMS})$ undergo contraction
because the neutrino energy loss rate exceeds the energy production
rate of C burning as shown in Figure \ref{fig:trho}.
After Si burning, the star with $M ({\rm ZAMS}) =$
15 M$_{\odot}$ exhibits a distinct behavior compared to other models
because shell Si burning is energetic.
However, the effects of shell Si burning are not the focus of this work
and will be discussed in future works.

More massive stars eject more materials but explode less frequently
\citep{1995Metic..30..325M}.
Considering the combined effects of ejected masses and event frequencies,
stars with $M ({\rm ZAMS}) =$ 25 M$_{\odot}$
are regarded as the most significant contributors to the chemical enrichment
of galaxies \citep{1978ApJ...225.1021W, 1995ApJS..101..181W}.
Therefore, we select the model of $M ({\rm ZAMS}) =$ 25 M$_{\odot}$  as a typical 
example for discussing stellar nucleosynthesis.

\begin{figure}[htb]
\centering
\begin{minipage}[c]{0.45\textwidth}
\includegraphics [width=80mm]{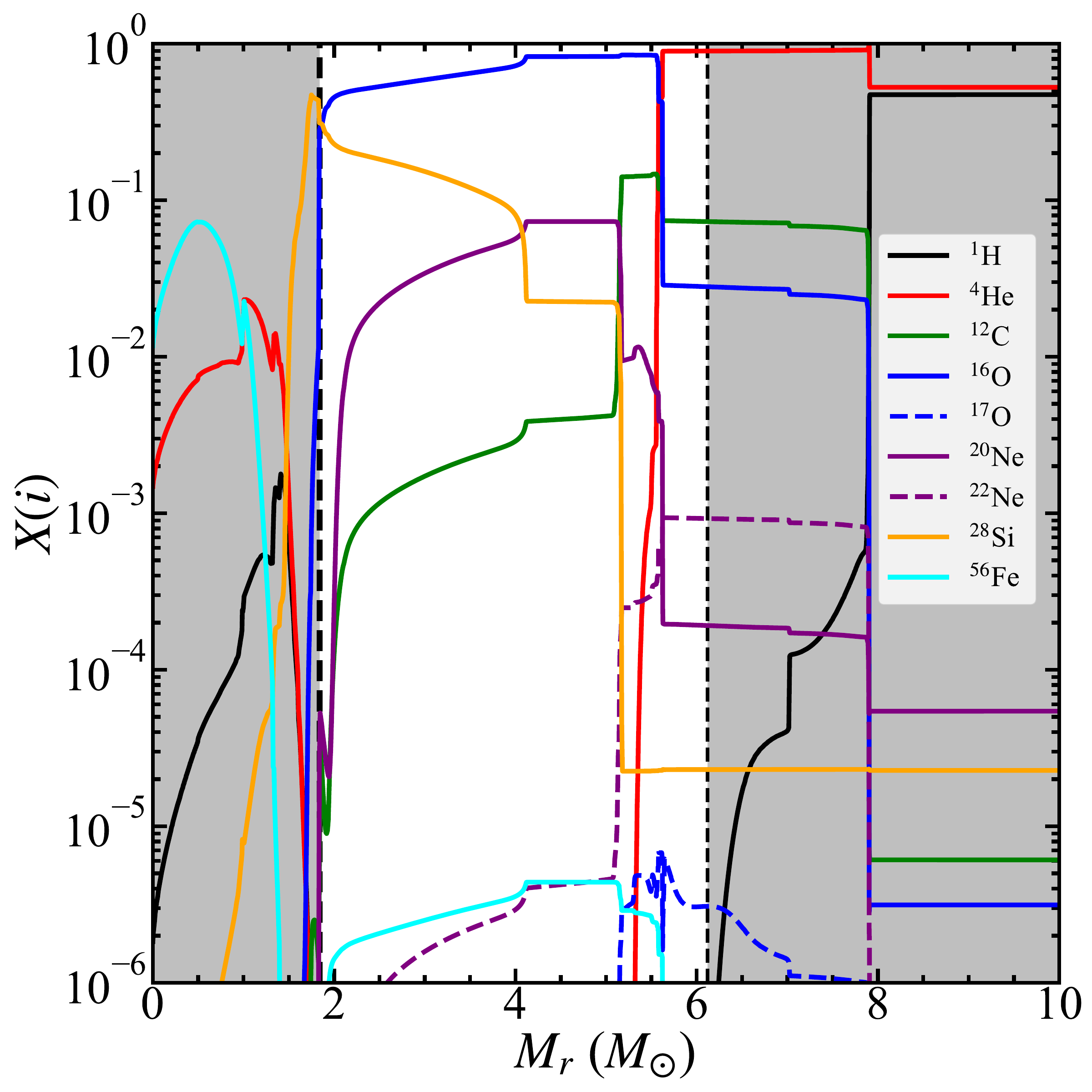}
%\centerline{(a) }
\end{minipage}%
\caption{The mass distribution of the main isotopes at $t=t_{\rm final}$ from MESA.
The two grey regions indicate where $M_r \le 1.84$ M$_\odot$
and $T\le 0.2$ GK, respectively.
\label{fig:25M_isos}}
\end{figure}

\begin{figure}[htb]
\centering
\begin{minipage}[c]{0.45\textwidth}
\includegraphics [width=80mm]{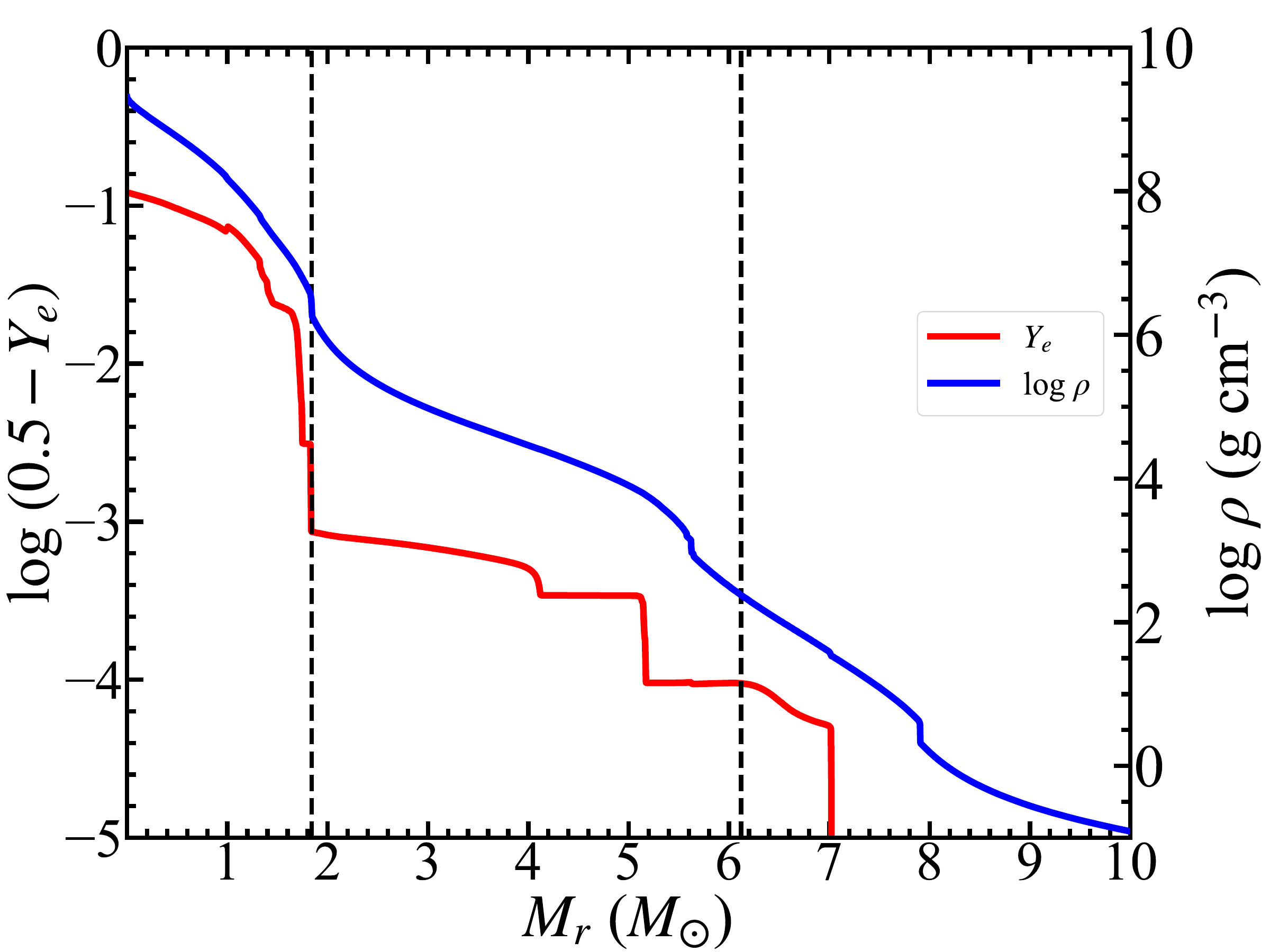}
%\centerline{(a) }
\end{minipage}%
\caption{The mass distribution of the $Y_e$ (red) and log $\rho$ (blue) at $t=t_{\rm final}$..
The two black lines indicate where $M_r = 1.84$ M$_\odot$
and $T = 0.2$ GK, respectively.
\label{fig:25M_yerho}}
\end{figure}

In Figure \ref{fig:25M_evo}, we present the Kippenhahn diagram for
the star with $M {\rm (ZAMS)}$ = 25 M$_{\odot}$,
tracking its evolution from H burning to Fe core collapse.
The central temperature reaches approximately 0.2 GK
at $\tau$ = $t_{\rm final}-t$ = 10$^{5.6}$ yr.
Here $t$ is the time from ZAMS and $t_{\rm final}$ denotes the time
at the final stage of evolution, which is defined as
the moment when the infall speed of the Fe core reaches 1000 km s$^{-1}$.
The orange line indicates the isotherm of $T =$ 0.2 GK.

The ws-process is assumed to occur interior to this isotherm.
After core He burning, this region extends to $M_r \sim$ 6.0 M$_\odot$,
and the He and CO core masses are 7.8 and 4.96 M$_{\odot}$, respectively.
C burning ignites off-center at $\tau = 10^{0.5}$ yr, nearly 3 years before the collapse.
After $\tau=10^{-3}$ yr (10 hours before collapse),
shell C burning merges with shell O burning at $M_r=$ 1.84 M$_{\odot}$,
marking the location of the highest energy generation rate
as indicated by the red line.
The inner part of this region is predicted to form a proto-neutron star (PNS),
while the outer layers are ejected. Therefore, 
this paper considers only the ws-process isotopes produced in the hatched region.

Figure \ref{fig:25M_isos} illustrates the mass distribution of
the main isotopes at $t = t_{\rm final}$. 
The ws-process region extends from the Si/O interface to the
bottom of He burning shell, primarily composed of
$^{16}$O, $^{28}$Si, $^{20}$Ne, $^{12}$C and $^{4}$He.
The mass fraction of iron-group elements ranges from $10^{-4}$ to $10^{-5}$.
Additionally, the neutron excess, expressed as $\eta = 1-2Y_e$,
of the ws-process region is observed to be 
$10^{-3}$ - $10^{-4}$, as shown in Figure \ref{fig:25M_yerho}.
In the interior to $M_r=1.84$ M$_\odot$, the neutron excess increases
rapidly toward the center, reaching $\eta \sim$ 0.2 in the center. 
This jump is primarily attributed to the reactions during O burning, 
including $^{16}$O($^{16}$O, n)$^{31}$S and the weak interactions
such as $^{30}$P(e$^+, \nu$)$^{30}$S, $^{33}$S(e$^-, \nu$)$^{33}$P,
$^{35}$Cl(e$^-, \nu$)$^{35}$S, and $^{37}$Ar(e$^-, \nu$)$^{37}$Cl
\citep{2002RvMP...74.1015W}.

\begin{figure}[htbp]
\centering
\begin{minipage}[c]{0.48\textwidth}
\includegraphics [width=85mm]{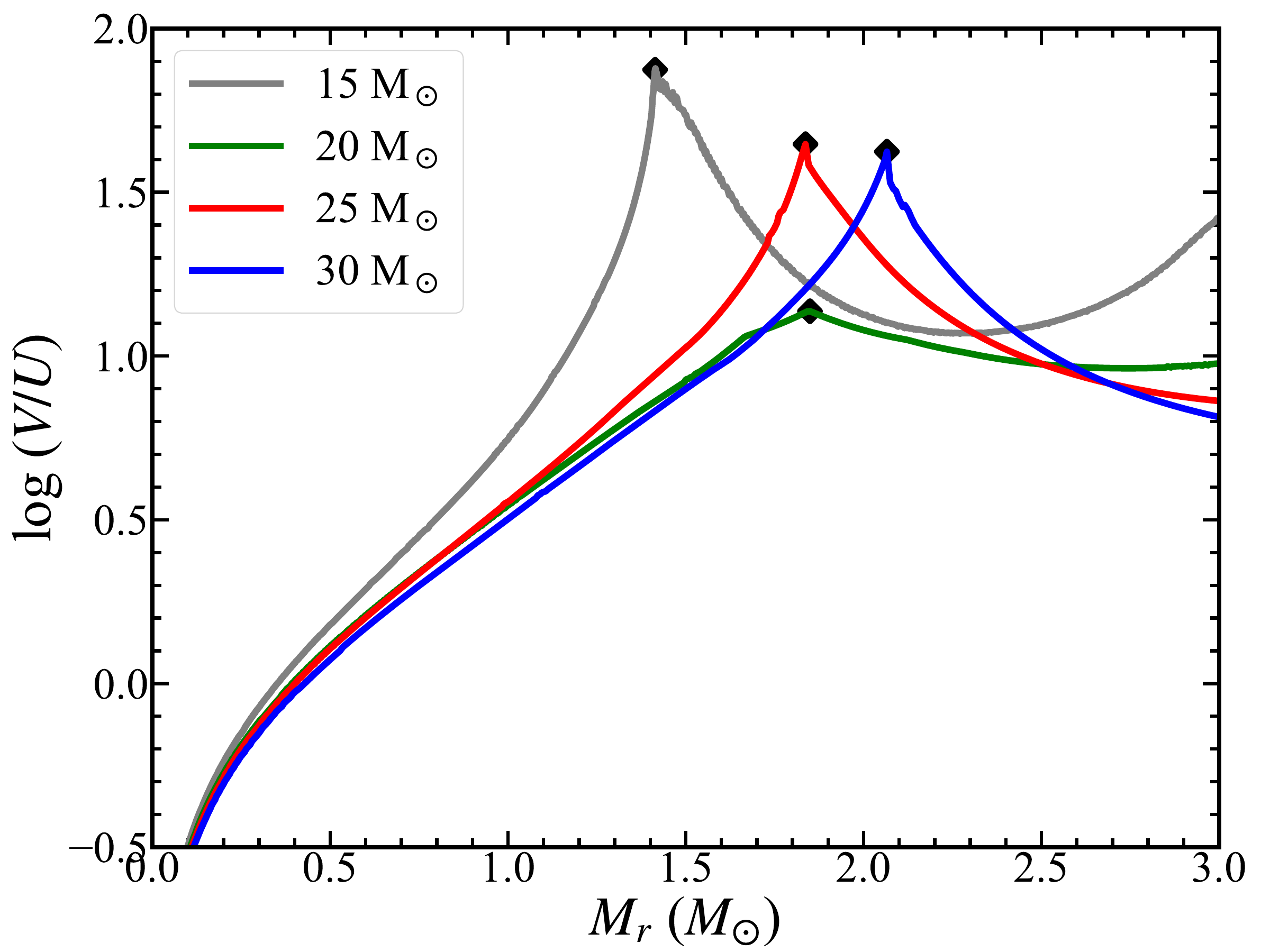}
%\centerline{(a) }
\end{minipage}%
\caption{The mass distribution of log ($V/U$) at $t=t_{\rm final}$.
The black diamonds are the location of $M_4$, which is defined at
the mass coordinate of specific entropy $s=4$ erg g$^{-1}$K$^{-1}$).
\label{fig:25M_uv}}
\end{figure}

\subsection{The Mass Cut}   \label{sec:exp_pre}

In Figure \ref{fig:25M_yerho}, prominent jumps in both density and 
$Y_e$ are observed near the mass coordinate $M_r = 1.84$ M$_\odot$.
This layer corresponds to the base of shell O burning, being 
the layer of the peak energy generation.
To measure the strength of shell burning, we use

\begin{equation} \label{equ:vu}
\frac{V}{U} = - \frac{{\rm dln} P}{{\rm dln} M_r} = \frac{G M_r^2}{4\pi r^4P}
\end{equation}                                                             
where $U$ and $V$ are defined in earlier studies
\citep{2015ses..book.....S, 1962PThPS..22....1H, 1980SSRv...25..155S, 2013sse..book.....K}.

As explained in detail in \citet{2025arXiv250211012X},
$U$ relates to the degree of the density jump 
and $V/U$ is the pressure gradient against $M_r$.
The mass coordinate where $V/U$ reaches its maximum is
represented as $M(V/U_{\rm max})$.
The relation between $V/U$ and the strength of shell burning
is straightforward. When shell O burning is more energetic,
it produces higher energy to prevent the contraction and even to cause
the expansion of outer layers, This makes the gradients of entropy and
pressure against $M_r$ (i.e., $V/U$) larger.

\begin{table}[htbp]  
\centering
\caption{The final mass, He core mass, CO core mass, and $M(V/U_{\rm max})$
for each model. The $M(\rm He)$ and $M(\rm CO)$ are defined at the layers
where $X$(H) and $X$(He) lower than 10$^{-4}$.}
\label{tab:cores}
\begin{tabular}{ccccc}
\toprule
$M(\rm ZAMS)$ &  $M(\rm final)$  &   $M(\rm He)$   &   $M(\rm CO)$   &   $M(V/U_{\rm max})$  \\
M$_{_\odot}$  &  M$_{_\odot}$    &   M$_{_\odot}$  &   M$_{_\odot}$  &   M$_{_\odot}$        \\
\midrule
    15        &  14.88  &   4.66          &   2.39          &   1.41     \\
    20        &  19.79  &   6.25          &   3.85          &   1.85     \\
    25        &  24.73  &   7.80          &   4.96          &   1.84     \\
    30        &  29.50  &   9.86          &   6.76          &   2.07     \\
\bottomrule
\end{tabular}
\end{table}

Figure \ref{fig:25M_uv} shows the distribution of log $V/U$
against $M_r$ at $\tau=t_{\rm final}$ for each model.
We note that $M(V/U_{\rm max})$ coincides with $M_4$, i.e.,
$M_r$ at a specific entropy of $s=4$ erg g$^{-1}$K$^{-1}$, 
being previously used for the mass cut which would divide 
the inner proto-neutron star (PNS) and the outer ejecta in
the explosion \citep{2007PhR...442..269W, 2010ApJ...724..341H, 
2016ApJ...821...38S, 2023ApJ...948..111F}.
In the present study, we adopt $M(V/U_{\rm max})$ as the mass cut,
because it is the location of the steepest gradients of pressure
and density \citep{2025arXiv250211012X}.
The core masses and $M(V/U_{\rm max})$
for our models are listed in Table \ref{tab:cores}.

\section{Nucleosynthesis and the effect of reaction rates} \label{sec:nuc}

\subsection{The Nucleosynthesis in the 25 M$_\odot$ model} \label{sec:one-zone}

In this section, we present the results of the post-process
nucleosynthesis and discuss the effects of updated reaction rates.

\begin{figure*}[htbp]
\centering
\begin{minipage}[c]{0.48\textwidth}
\centerline{(a) Default Rates}
\includegraphics [width=85mm]{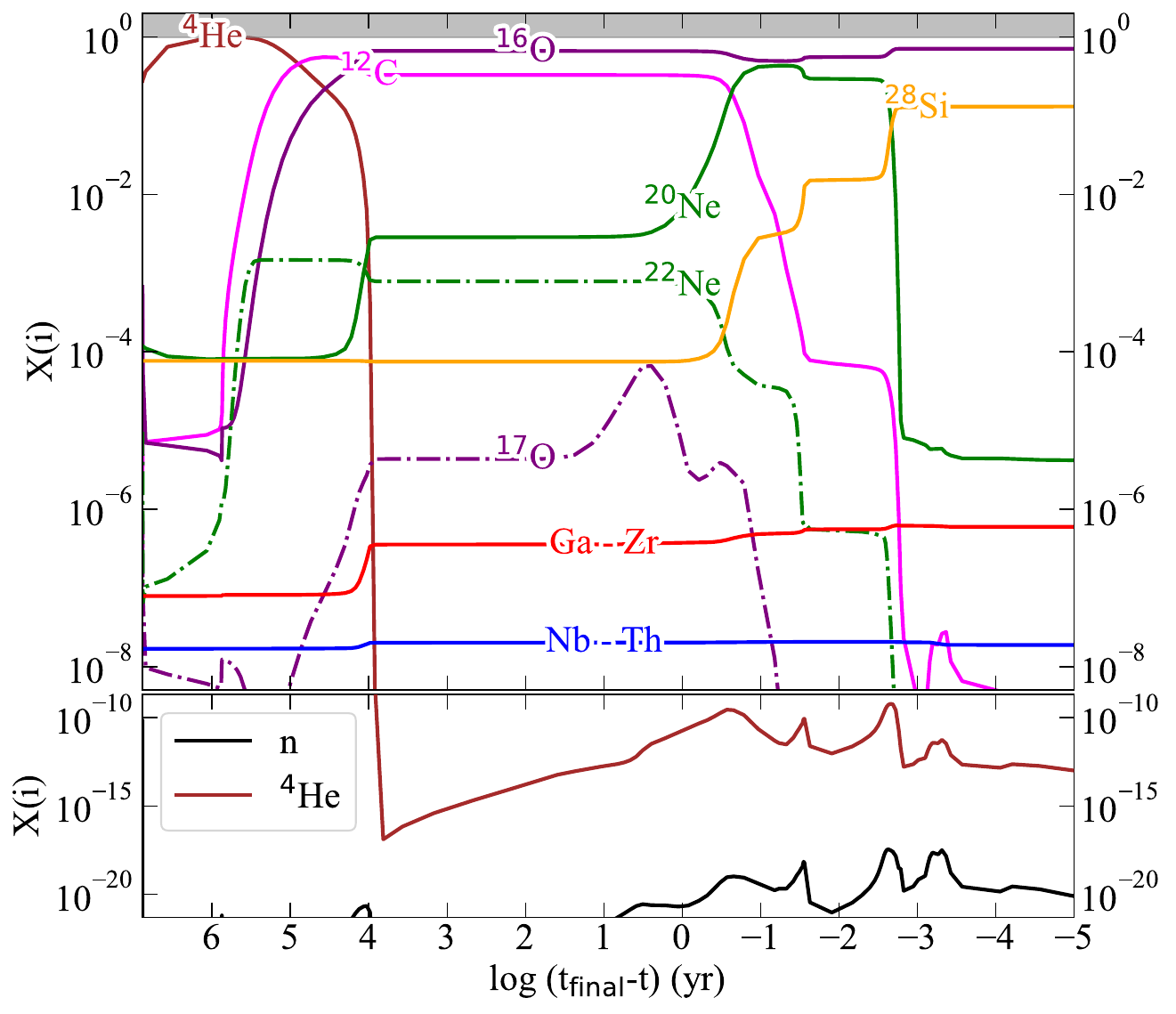}
\end{minipage}%
\begin{minipage}[c]{0.48\textwidth}
\centerline{(b) New $^{22}$Ne+$\alpha$ Rates}
\includegraphics [width=85mm]{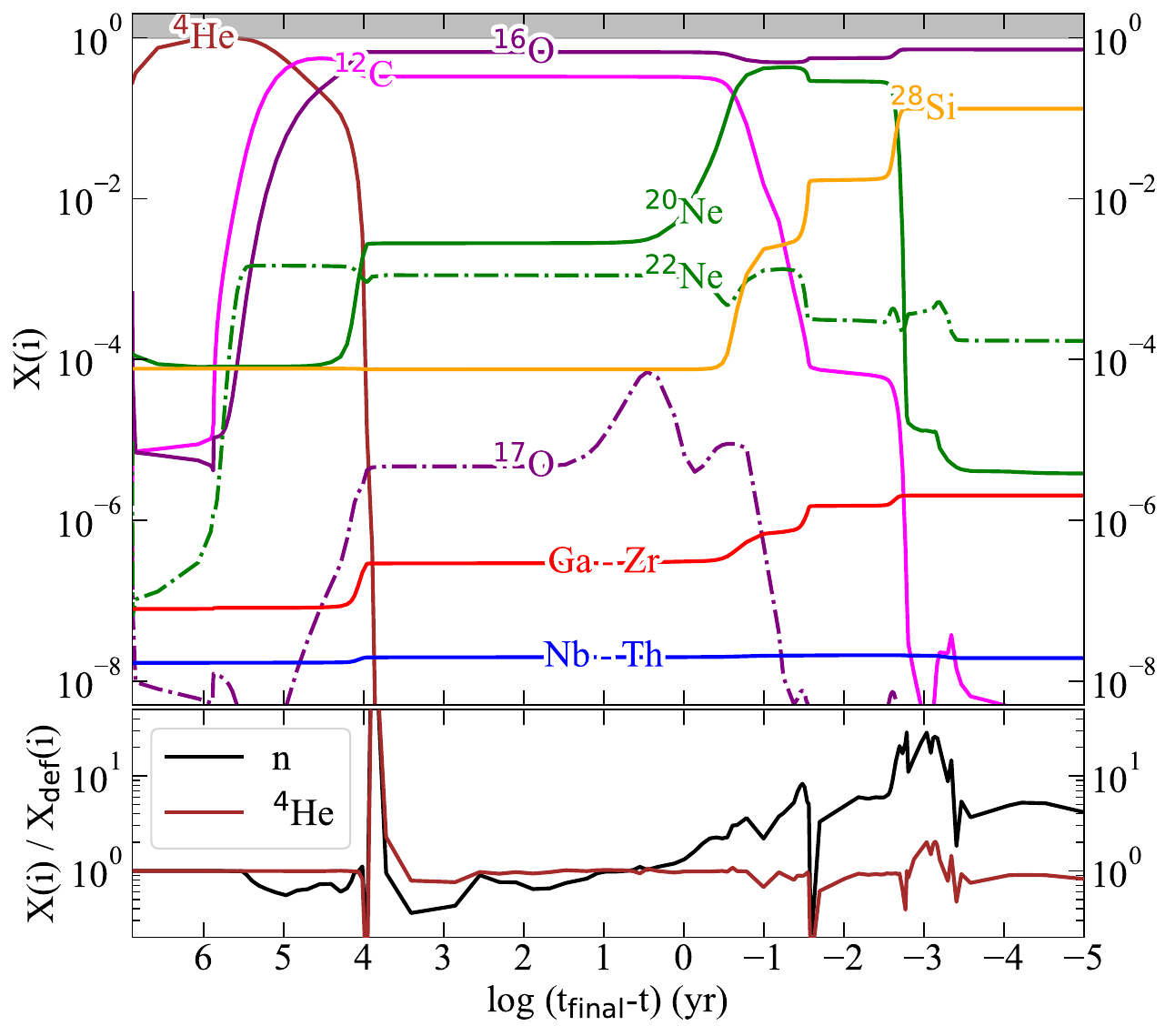}
\end{minipage}%

\begin{minipage}[c]{0.48\textwidth}
\centerline{(c) New $^{17}$O+$\alpha$ Rates }
\includegraphics [width=85mm]{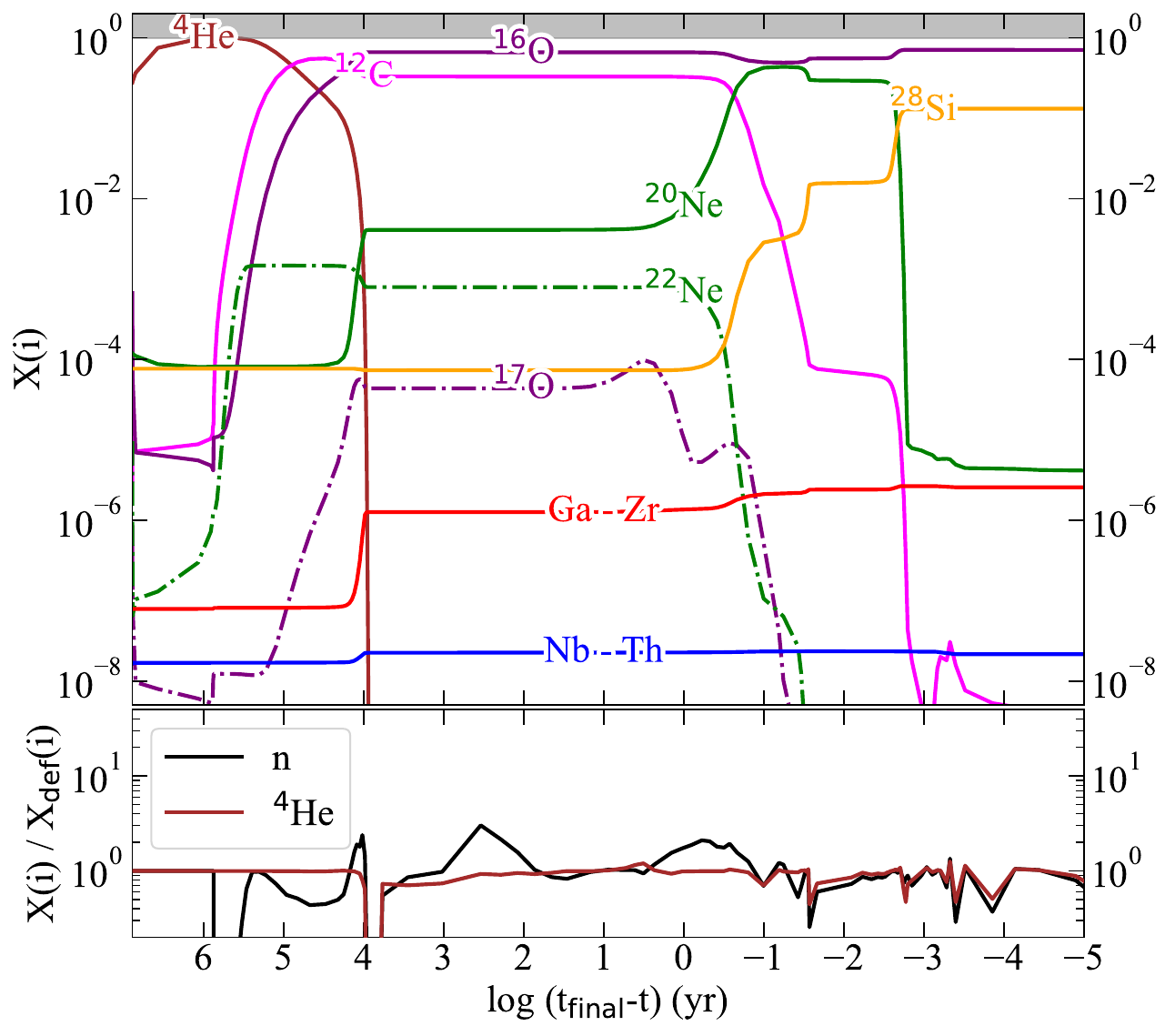}
\end{minipage}%
\begin{minipage}[c]{0.48\textwidth}
\centerline{(d) All New Rates}
\includegraphics [width=85mm]{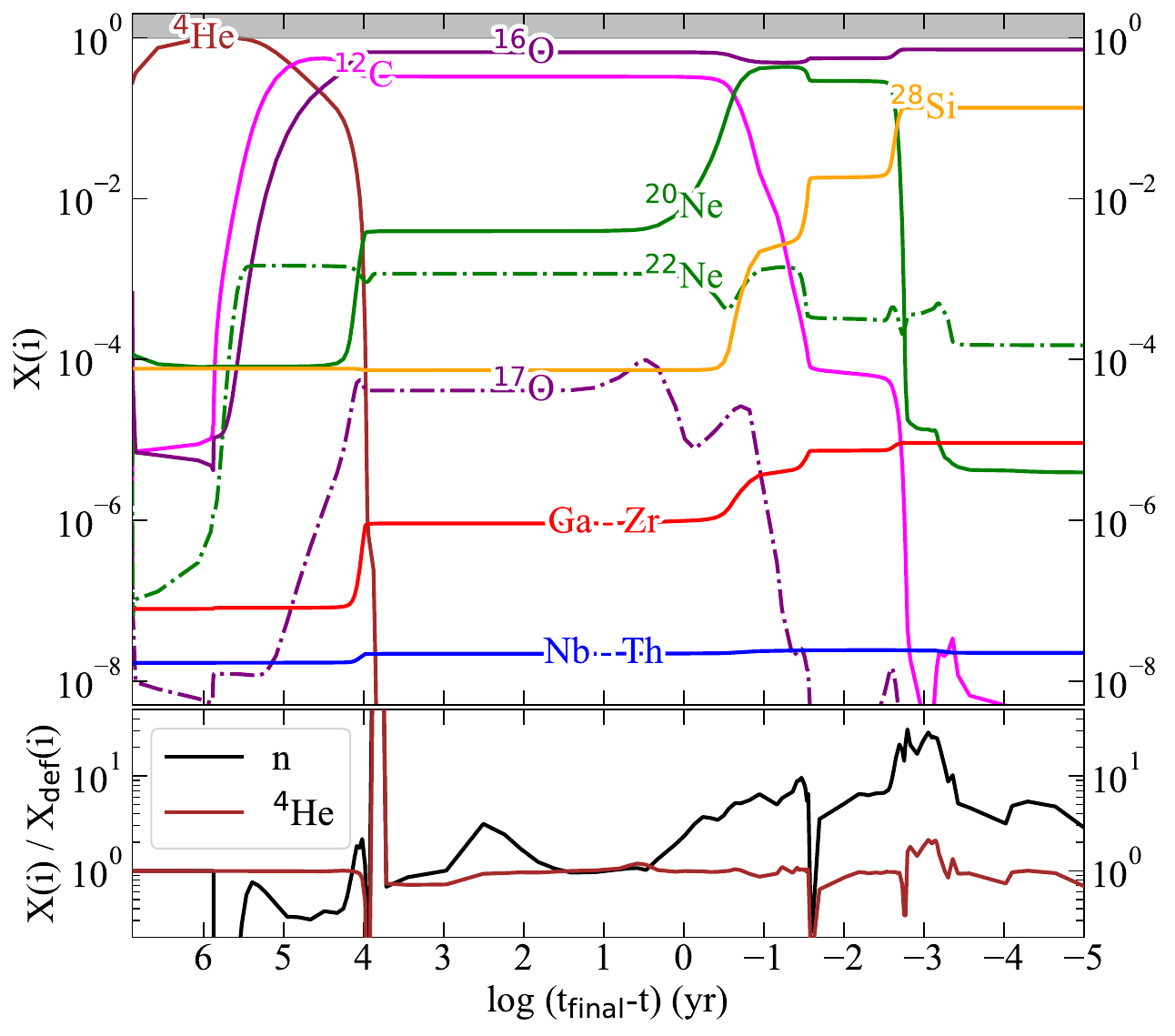}
\end{minipage}%
\caption{
The chemical evolution in the OSi shell ($M_r=2.3$ M$_\odot$).
In the top panels of (a) - (d), changes in the mass fractions $X$(i) 
of $^4$He, $^{12}$C, $^{16}$O, and $^{20}$Ne show the burning stages.
$X$(Ga - Zr) and $X$(Nb - Th) are the cumulative mass fractions
of the s-process isotopes from Ga to Zr ($Z = 31 - 40$) and Nb to Th
($Z>40$), respectively. 
In the bottom panel, the changes in $X_{\rm def}$(i) of neutrons and
$^{4}$He for the default rates are shown in (a).
In (b) - (d) shown are the changes in the ratios $X$(i)/$X_{\rm def}$(i)
between the ``new'' and default rates for n and $^{4}$He.
\label{fig:25M_evo}}
\end{figure*}

\begin{figure}[htb]
\centering
\begin{minipage}[c]{0.48\textwidth}
\includegraphics [width=80mm]{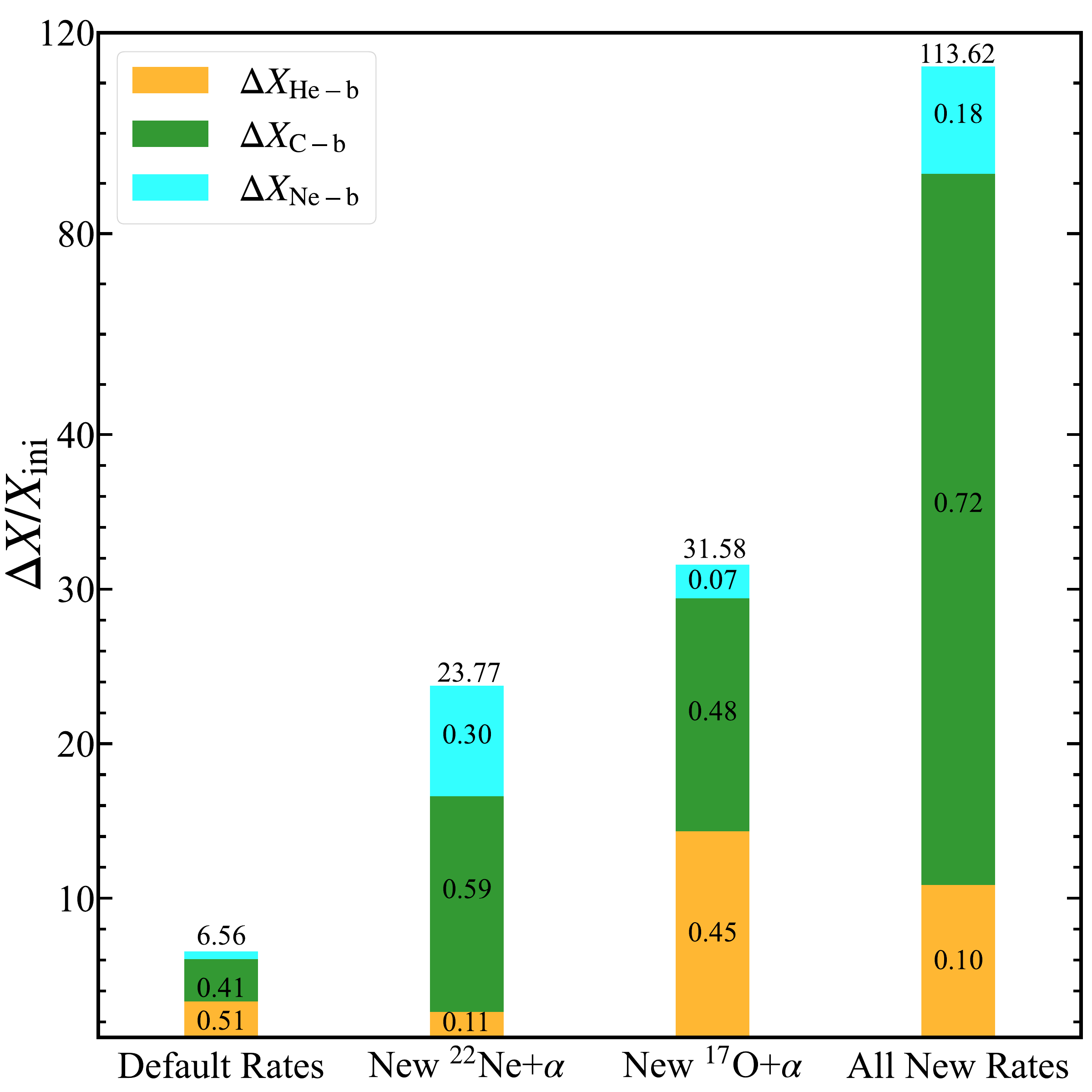}
%\centerline{(a) }
\end{minipage}%
\caption{The total mass fraction of the ws-process isotopes
from $A=31-40$ after He, C, and Ne burning.
All values are normalized by the initial abundance,
$X_{\rm ini}=7.87 \times10^{-8}$.
The total value is written above each pillar and the
fractional contribution of each burning
stage is written in the center of the box.
\label{fig:burn_enhance}}
\end{figure}

We selected a zone at $M_r$ = 2.3 M$_\odot$ as a representative
example to reveal the change in the mass fraction $X$(i) of isotope i
occurring during each burning stage after updating the ($\alpha$, n) reaction rates. 
$X$(Ga-Zr) and $X$(Nb-Th) are the cumulative mass fractions
of isotopes from Ga to Zr ($Z = 31 - 40$) and Nb to Th ($X>40$), respectively. 
The reaction rates recipes used in Figure \ref{fig:25M_evo} (a - d)
correspond to the cases 1 - 4 listed in Table \ref{tab:recipes}.
The chemical evolution of main isotopes is depicted in the top panel of
Figure \ref{fig:25M_evo}. The changes in $X$(n) and $^{4}$He for the default
rates are displayed in the bottom panel of Figure \ref{fig:25M_evo} (a),
while in Figure \ref{fig:25M_evo} (b) - (d), the value is normalized by
$X_{\rm def}$(i) in Figure \ref{fig:25M_evo} (a) to stress the effect of
these reaction rates.
Table \ref{tab:mass-frac} lists the total mass fraction of the ``Ga - Zr" elements
in the initial abundance ($X_{\rm ini}$), after He burning ($X_{\rm He-b}$),
C burning ($X_{\rm C-b}$), and Ne burning ($X_{\rm Ne-b}$). 
The data is visualized in Figure \ref{fig:burn_enhance}
by the ratios of $\Delta X$ to $X_{\rm ini}$, where $\Delta X$ is the change in the
mass fraction in each burning stage.

Overall, these new reaction rates significantly alter the production of 
the ``Ga - Zr" elements rather than the ``Nb - Th" elements.
We thus focus on the ``Ga - Zr" elements in this section.
The initial value of $X$(Ga-Zr) is $X_{\rm ini} = 7.87 \times10^{-8}$.
Enhancements of $X$(Ga-Zr) are observed 4 times
at the end of He burning, the beginning of C burning, the end of
C burning and Ne burning, respectively. They coincide with the
neutron peaks and $^{4}$He production in Figure \ref{fig:25M_evo}.

After Ne burning, the total enhancement of $X$(Ga-Zr) is estimated
by a ratio of ($X_{\rm Ne-b}$-$X_{\rm ini}$)/$X_{\rm ini}$, where
$X_{\rm Ne-b}$ and $X_{\rm ini}$ are listed in Table \ref{tab:mass-frac}.
Compared with $X_{\rm ini}$, $X$(Ga-Zr) increases by a factor of
6.56, 23.77, 31.58 and 113.62 for (a) - (b), respectively.
The forthcoming O burning will not enhance but rather reduce their abundance
because of more destruction at high temperatures \citep{2009ApJ...702.1068T}.

\begin{table}[htbp] 
\centering
\caption{The total mass fractions of the ws-process isotopes of $Z=31-40$ in the
initial abundance, after He burning, C burning and Ne burning stages, respectively,
in Figure \ref{fig:25M_isos}. }
\label{tab:mass-frac}
\begin{tabular}{ccccc}
\toprule
Figure   &  $X_{\rm ini}$  &  $X_{\rm He-b}$  &  $X_{\rm C-b}$   &   $X_{\rm Ne-b}$  \\
\midrule
    a   &  7.87E-08  &   3.42E-07   &  5.55E-07    &   5.95E-07    \\
    b   &  7.87E-08  &   2.87E-07   &  1.38E-06    &   1.95E-06    \\
    c   &  7.87E-08  &   1.21E-06   &  2.39E-06    &   2.56E-06    \\
    d   &  7.87E-08  &   9.34E-07   &  7.34E-06    &   9.02E-06    \\
\bottomrule
\end{tabular}
\end{table}

When the default rates are used as in Figure \ref{fig:25M_evo} (a),
the ``Ga - Zr" elements are mainly synthesized during the He (51\%) and
C (41\%) burning stages (see Figure \ref{fig:burn_enhance}).
Only 8\% of them are synthesized during the Ne burning stage because the
main neutron source isotope $^{22}$Ne is almost exhausted.

With the new $^{17}$O+$\alpha$ reaction rates in Figure \ref{fig:25M_evo} (c), 
more than 93\% of the ``Ga - Zr" elements are synthesized during He and C
burning stages, being similar to (a). The final $X$ (Ga - Zr) is
enhanced by a factor of 4.81 compared with default rates,
Because both new $^{17}$O($\alpha$, n)$^{20}$Ne and
$^{17}$O($\alpha$, $\gamma$)$^{21}$Ne reaction rates are lower
than the default ones at temperatures below 0.7 GK (see Figure \ref{fig:rate1}),
$X$($^{17}$O) reaches a higher level at the end of He burning with the new rates.
When the temperature exceeds 0.7 GK (see Section \ref{sec:diff_rate}),
the ratio of ($\alpha$, n)/($\alpha$, $\gamma$)
increases. Therefore, the new $^{17}$O+$\alpha$ reaction rates
significantly enhance the production of the ``Ga - Zr" elements
at all stages, though only slightly alter their contribution percentages.

Comparing Figure \ref{fig:25M_evo} (a) and (b), the productions of the 
``Ga - Zr" elements are enhanced by a factor of 23.8 by using
new $^{22}$Ne+$\alpha$ reaction rates, 
This is smaller than the increase in using the new $^{17}$O+$\alpha$ reaction rates. 
Since the new $^{22}$Ne+$\alpha$ rates are smaller than those in REACLIB 
(see Figure \ref{fig:rate1}), $^{22}$Ne is not exhausted until the core collapse. 
The ($\alpha$, n)/($\alpha, \gamma$) ratio of the new rates is 10 times higher
than that of the default ones when the temperature exceeds 1.5 GK.
A significant neutron rise is observed from six months before the explosion.
As a result, almost 89\% of the ``Ga - Zr" elements are synthesized during
C and Ne burning.

In Figure \ref{fig:25M_evo} (d), both the new $^{17}$O+$\alpha$ and
new $^{22}$Ne+$\alpha$ reaction rates are updated. The production of the ``Ga - Zr"
elements are enhanced by more than one order of magnitude. But the contributions
of He and Ne burning are only 10\% and 18\% and most of the ``Ga - Zr" elements 
are synthesized during the C burning stage, which should alter the isotope
composition of the ``Ga - Zr" elements. Comparing (a, c) with (b, d),
we can also note whether Ne burning contributes to the ws-process significantly,
depending on the $^{22}$Ne($\alpha, \gamma$)$^{25}$Mg reaction rate.

\begin{figure}[htbp]
\centering
\begin{minipage}[c]{0.48\textwidth}
%\centerline{(a) Default Rates}
\includegraphics [width=85mm]{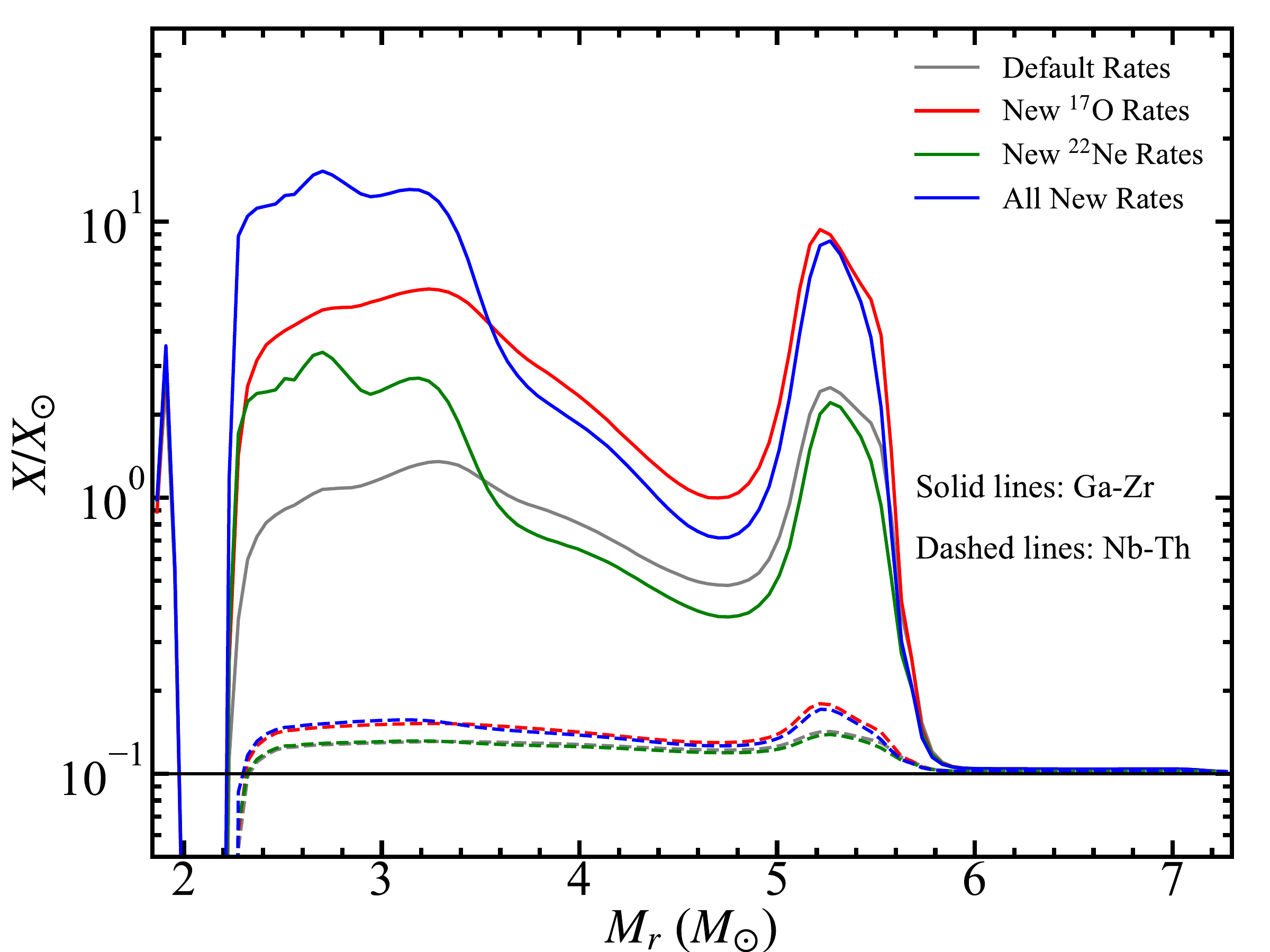}
\end{minipage}%
\caption{The abundance distributions of the ws-process elements at
$M_r$ = 1.84 - 7.3 M$_\odot$ in the He core of the
$M(\rm ZAMS)=25$ M$_\odot$ star.
The solid lines symbolize the ``Ga - Zr" elements,
while the dashed lines symbolize the heavier.
\label{fig:pre2}}
\end{figure}

\begin{figure}[htbp]
\centering
\begin{minipage}[c]{0.48\textwidth}
%\centerline{(a) Default Rates}
\includegraphics [width=85mm]{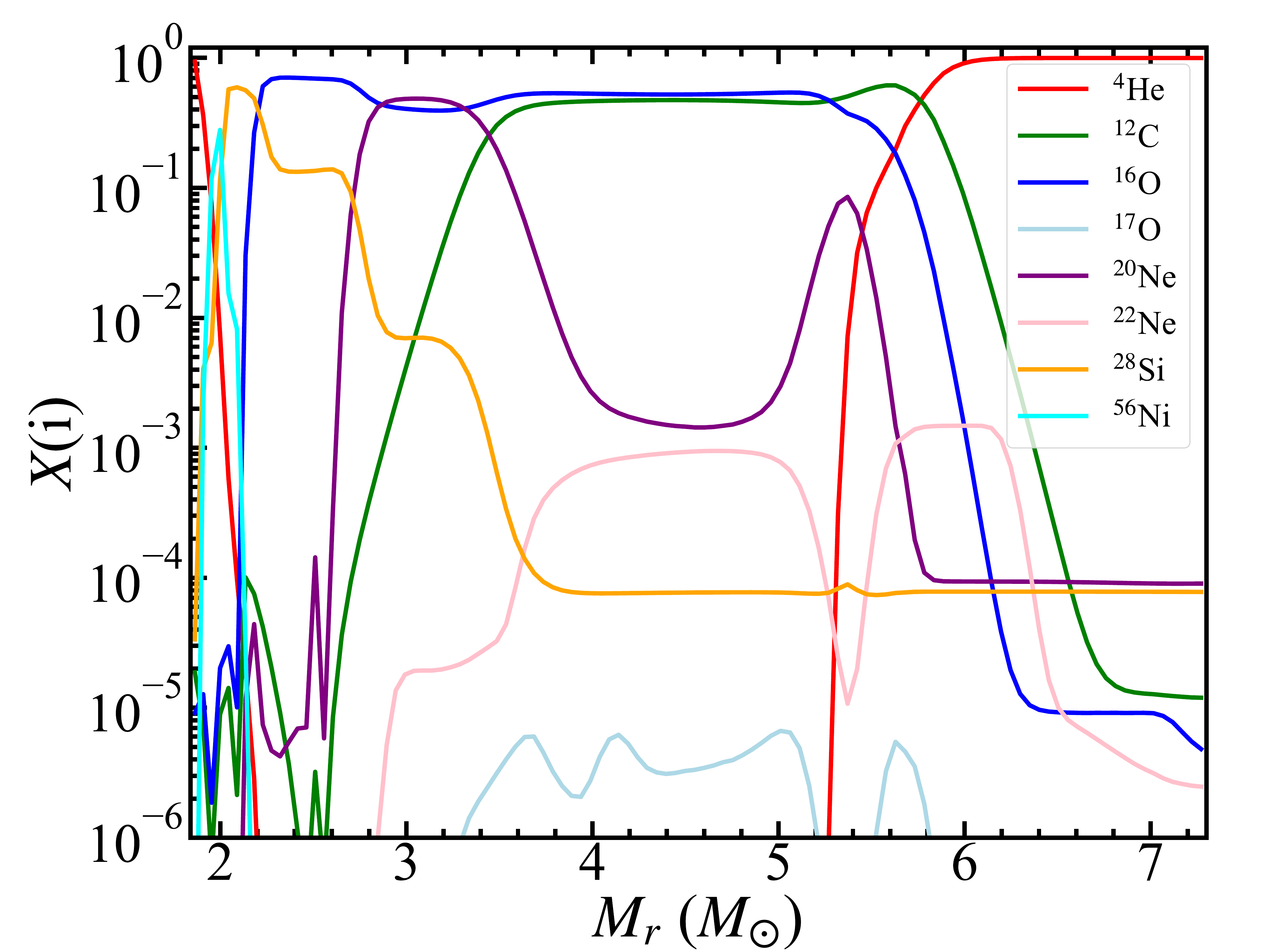}
\end{minipage}%
\caption{The distribution of the mass fractions of the main isotopes 
at $M_r$ = 1.84 - 7.3 M$_\odot$ obtained 
from the post-processing calculation.
Note that the mixing is not taken into consideration in the
post-processing calculation because WinNet is a one-zone code.
The comparison between this Figure and Figure \ref{fig:25M_isos}
is made in Section \ref{sec:mix}.
\label{fig:dist_abun}}
\end{figure}

\begin{figure*}[htbp]
\centering
\begin{minipage}[c]{0.8\textwidth}
%\centerline{(a) Default Rates}
\includegraphics [width=130mm]{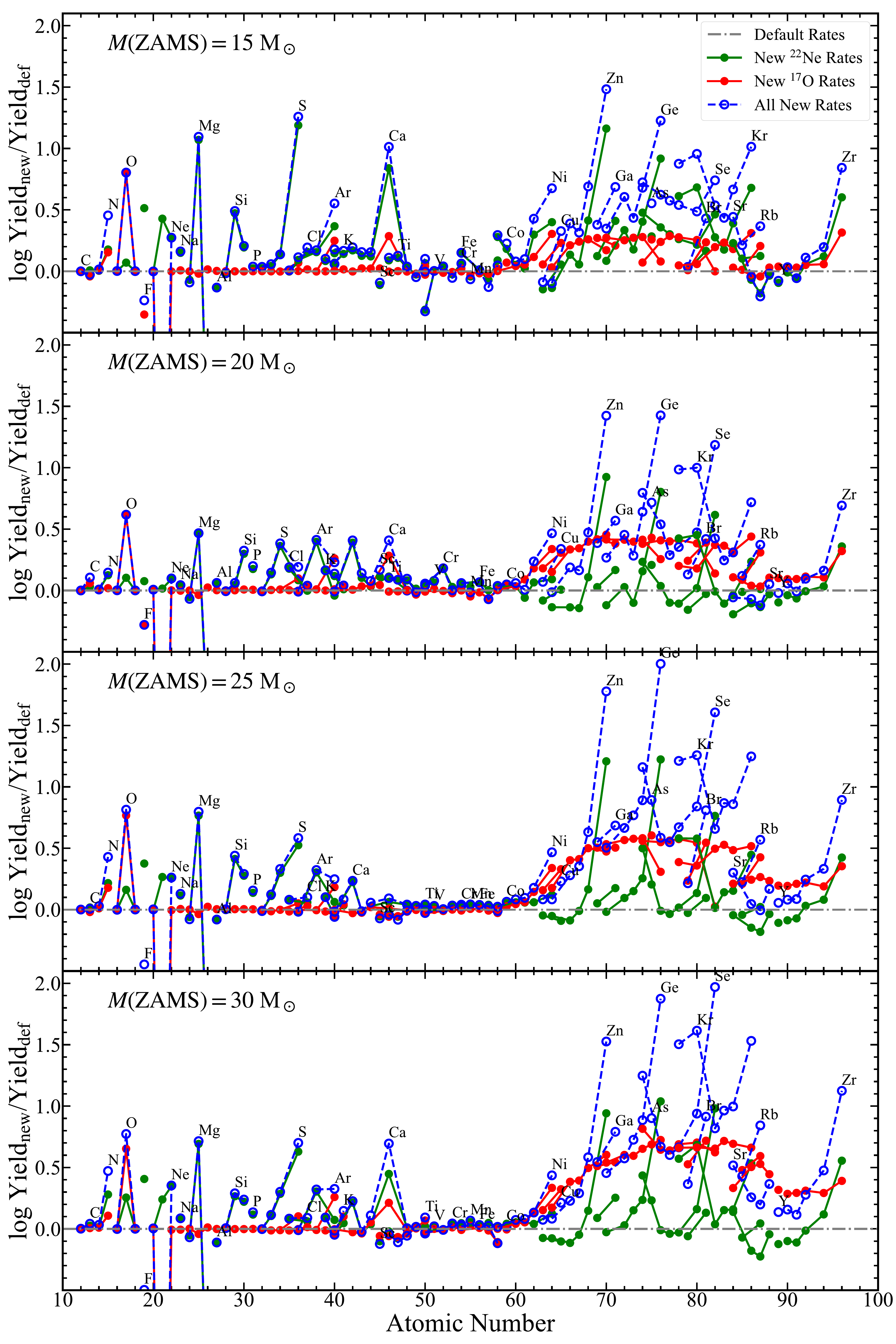}
\end{minipage}%
\caption{The ratios of isotopic yields (from C to Zr) between new rates and
default rates for $M(\rm ZAMS)=$ 15, 20, 25 and 30 M$_\odot$, respectively.
\label{fig:yields}}
\end{figure*}

\begin{figure*}[htbp]
\centering
\begin{minipage}[c]{0.8\textwidth}
%\centerline{(a) Default Rates}
\includegraphics [width=130mm]{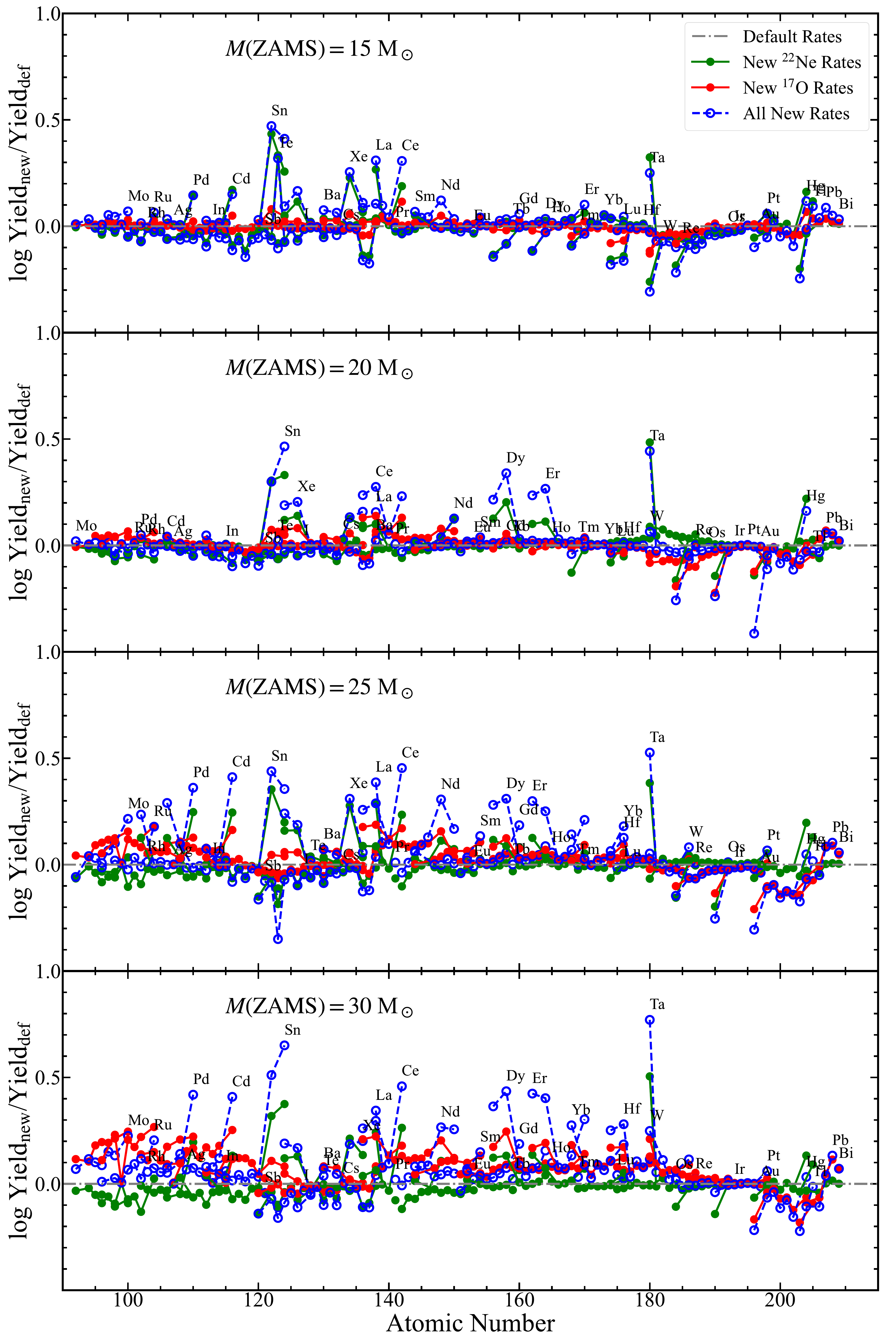}
\end{minipage}%
\caption{Same as Figure \ref{fig:yields}, but for elements from Mo to Bi.
\label{fig:yields2}}
\end{figure*}

Figure \ref{fig:pre2} shows the abundance distributions of s-process elements.
We observe two distinct bumps in the abundance of both ``Ga - Zr''
and ``Nb - Th'' elements in the region of $M_r$ = 2.2 - 5.9 M$_\odot$.
The first bump, located at $M_r$ = 2.0 - 3.6 M$_\odot$,
corresponds to the C, Ne, and O burning shells,
while the second bump, found at $M_r$ = 5.0 - 5.9 M$_\odot$,
is associated with shell He burning.
Between these two bumps, $X$ (Ga-Zr)
decreases due to low $\alpha$ production in the unburned regions.

In Figure \ref{fig:dist_abun},
the distribution of $^{21}$Ne is similar to the ``Ga - Zr'' elements,
while $^{22}$Ne displays an opposite trend.
Compared to the unburned shells, there are more $^4$He produced in
the burning shells, which can consume $^{22}$Ne and release
more neutrons.
New $^{22}$Ne+$\alpha$ reaction rates enhance
$X$ (Ga-Zr) only in the first bump,
whereas the new $^{17}$O+$\alpha$ reaction rates positively affect
$X$ (Ga-Zr) across all the ws-process regions.
Additionally, the new $^{17}$O+$\alpha$ reaction rates increase
$X$ (Nb - Th) by 50\%,
unlike the $^{22}$Ne+$\alpha$ reaction rates since these elements
are produced during He burning stage.
Notably, the sharp peak observed at $M_r$ = 1.84 - 2.0
M$_\odot$ remains unaffected by both the $^{22}$Ne+$\alpha$
and $^{17}$O+$\alpha$ reaction rates, as these ``Ga - Zr" elements
are generated through the NSE process.

In this section, we follow the variation of the ws-process isotopes
throughout the stellar evolution history and their mass distribution
at the final stage for various reaction recipes. We find both new
$^{22}$Ne+$\alpha$ and $^{17}$O+$\alpha$ reaction rates increase
the production of the ws-process isotopes. 

\noindent (1) The new $^{17}$O+$\alpha$ reaction rates only increase
neutron density by $\sim$3 times during He and C burning stages.
On the contrary, the new $^{22}$Ne+$\alpha$ reaction rates increase the
neutron density by several tens of times during the C and Ne burning.

\noindent (2) The new $^{17}$O+$\alpha$ reaction rates doesn't vary
the contribution in each burning stage.
On the contrary, new $^{22}$Ne+$\alpha$ reaction rates significantly increase
contribution in C and Ne burning stages, but decrease that in He burning stage.

\noindent (3) Before the explosion, the ws-process isotopes are
primarily concentrated in the burning shells,
with their abundances decreasing in the outer layers of the CO core.
This is because there is no C burning in the outer layers of the CO core
so that little $^4$He is released.

\subsection{The Integrated Yields} \label{sec:star}

As indicated in Figure \ref{fig:25M_evo}, the ws-process isotopes
produced between the mass cut and the top of He burning shell
may contribute to the overall enhancement.
We integrate all the isotopes in this region rather than the
entire star. We assume that the modifications 
of the ws-process yields in explosive nucleosynthesis can be
ignored and that all the radioactive isotopes decay into stable
ones after the explosion.
To investigate the sensitivity on the reaction rates, 
such approximations are reasonable.

In Figure \ref{fig:yields} and \ref{fig:yields2},
we show the ratios between the yields with the new reaction rates
(Yield$_{\rm new}$) and those with default reaction rates
(Yield$_{\rm def}$) for $M(\rm ZAMS)=$ 15, 20, 25 and 30
M$_\odot$, respectively.

\begin{figure*}[htbp!]
\centering
\begin{minipage}[c]{0.45\textwidth}
\includegraphics [width=80mm]{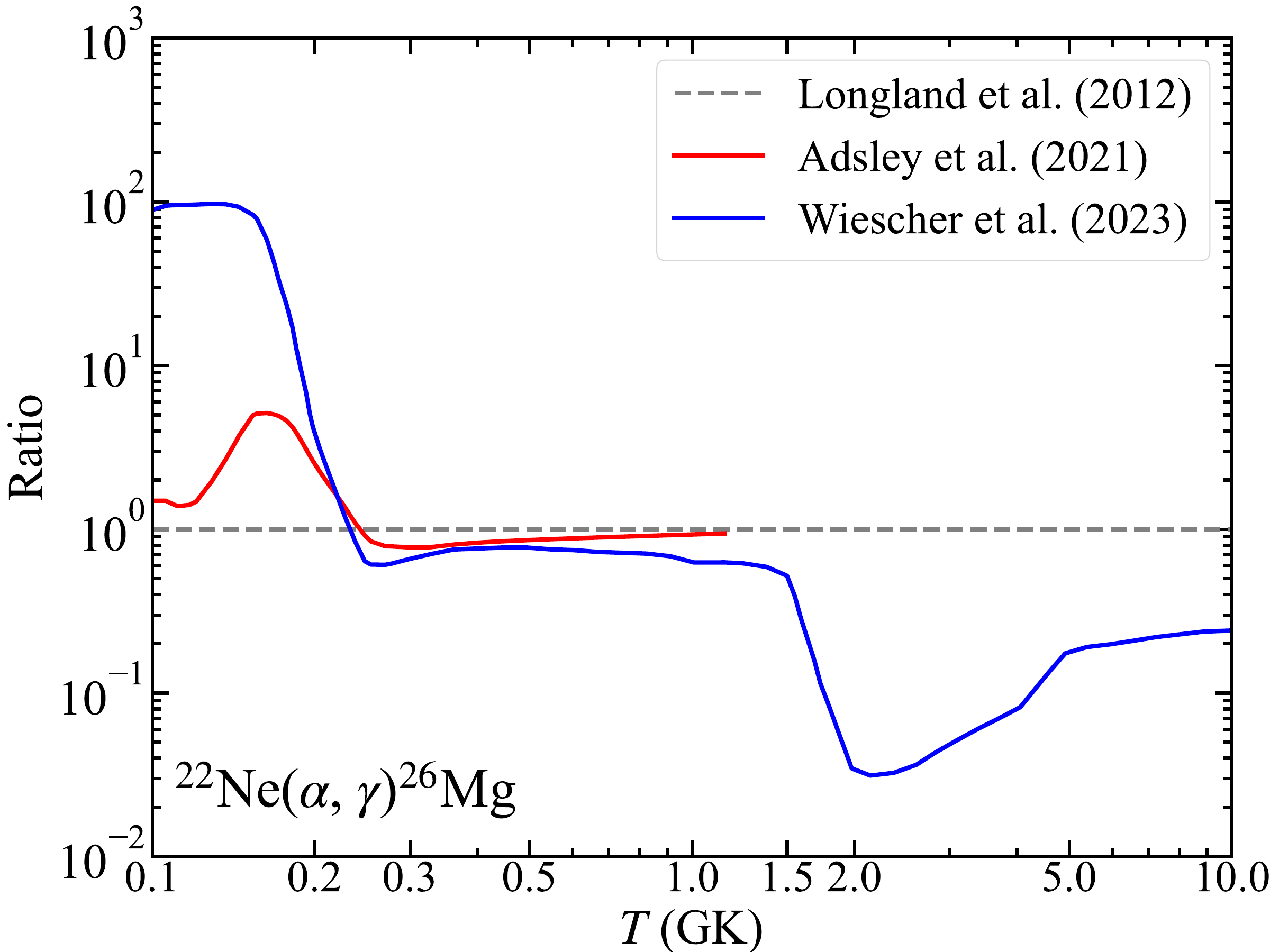}
\centerline{(a)}
\end{minipage}%
\begin{minipage}[c]{0.45\textwidth}
\includegraphics [width=80mm]{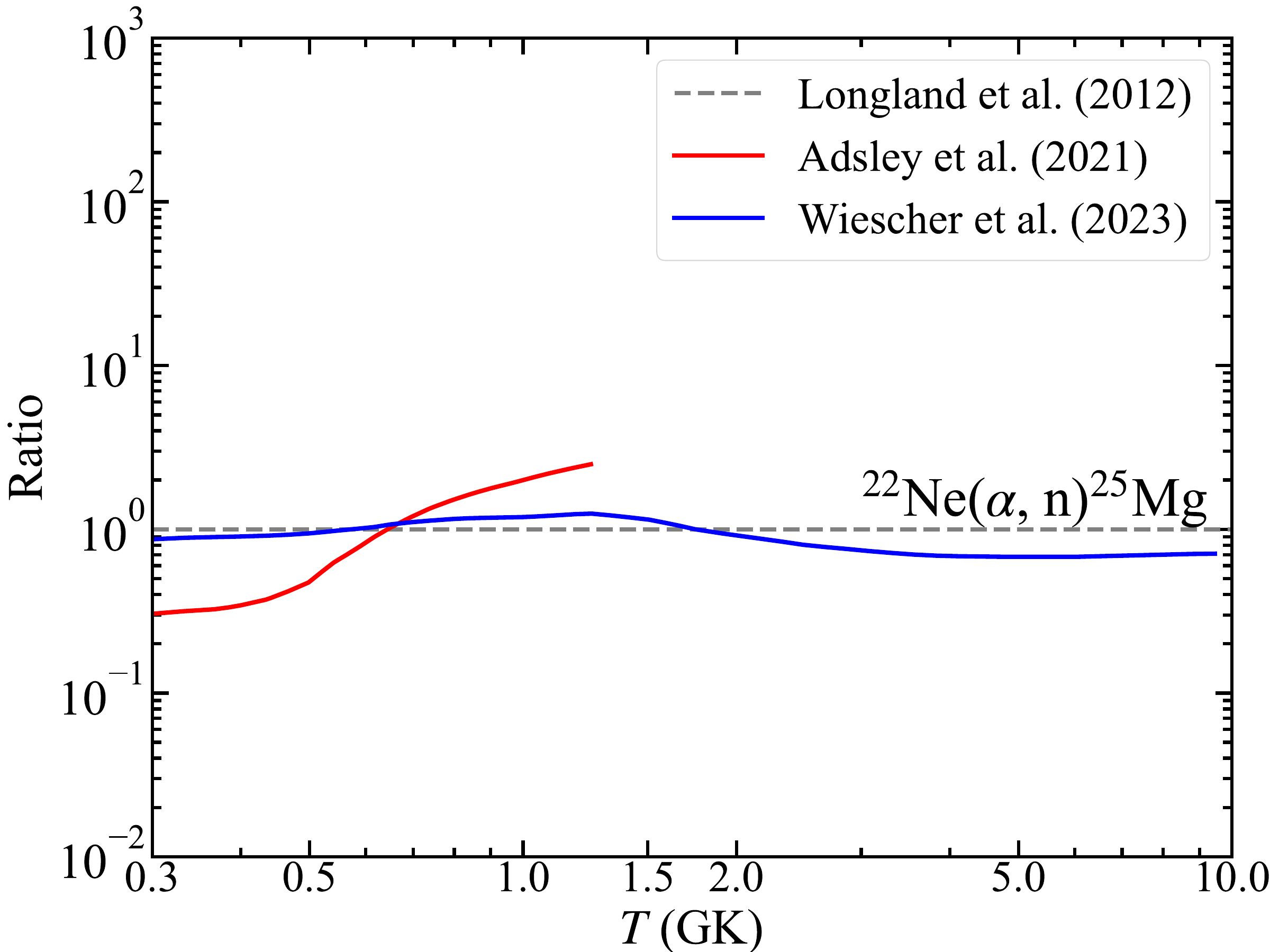}
\centerline{(b)}
\end{minipage}%

\begin{minipage}[c]{0.45\textwidth}
\includegraphics [width=80mm]{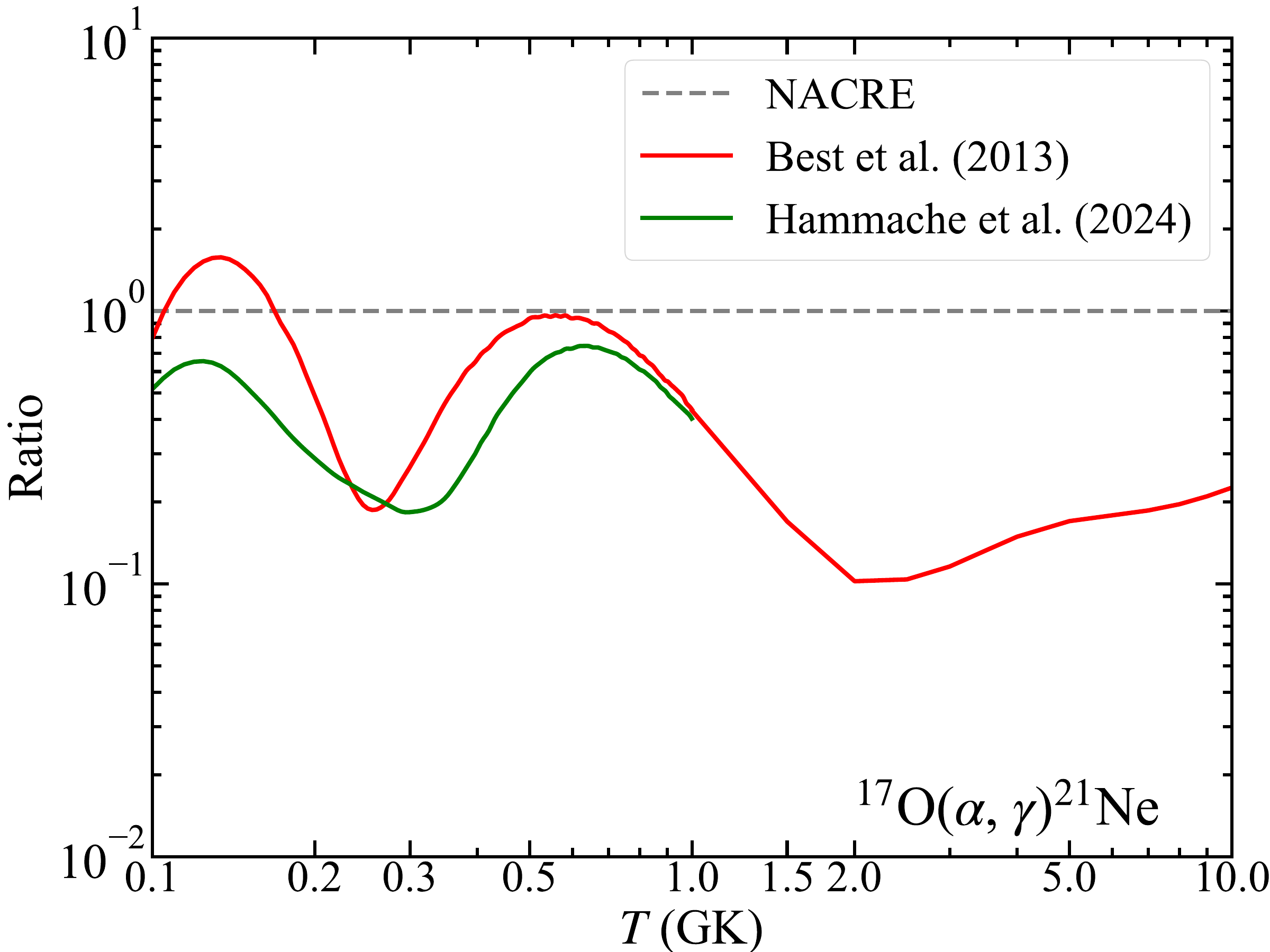}
\centerline{(c)}
\end{minipage}%
\begin{minipage}[c]{0.45\textwidth}
\includegraphics [width=80mm]{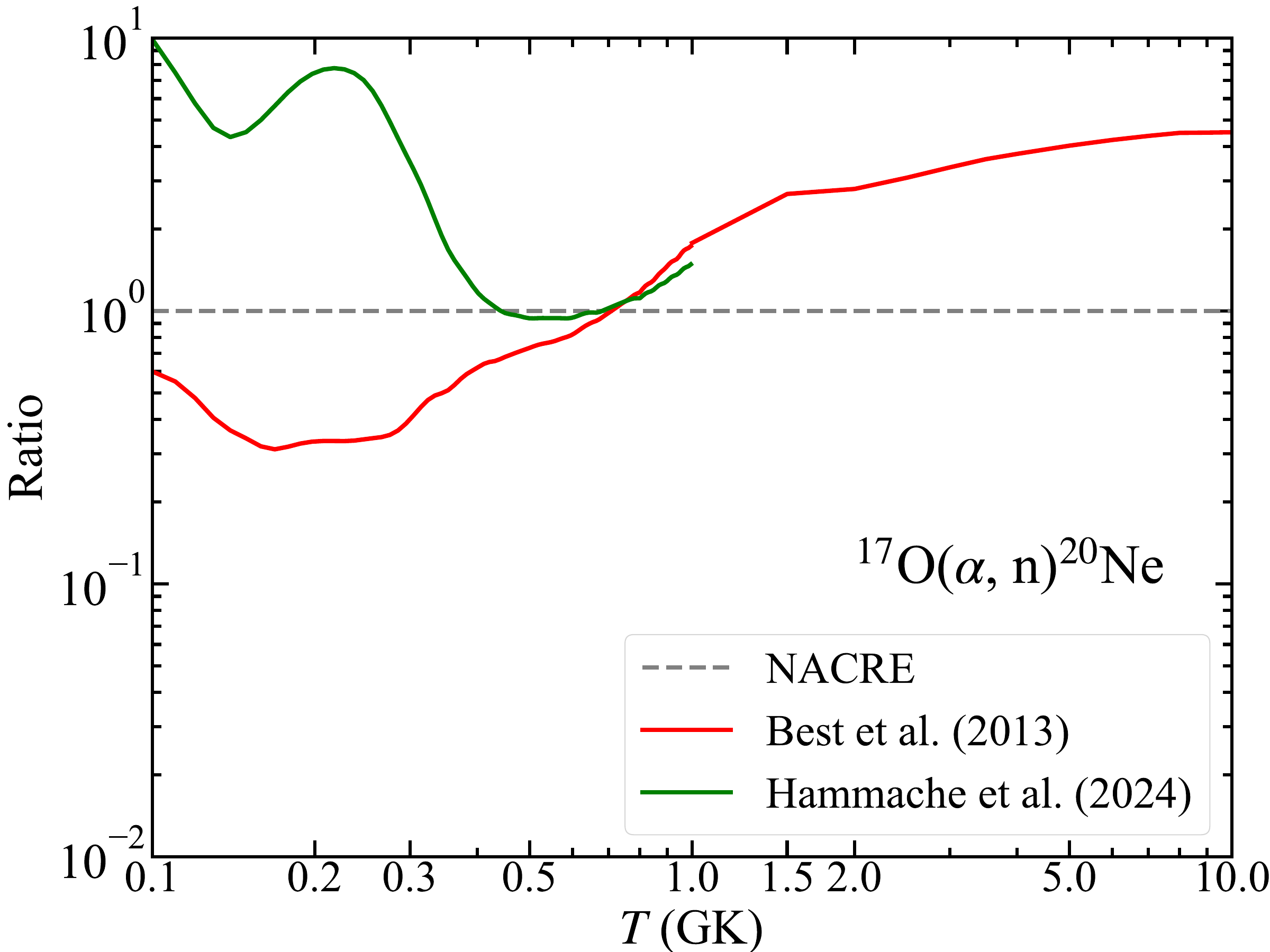}
\centerline{(d)}
\end{minipage}%
\caption{Top: $^{22}$Ne($\alpha$, $\gamma$)$^{26}$Mg and $^{22}$Ne($\alpha$, n)$^{25}$Mg
reaction rates from different references. The gray dashed line shows the default rates
in JINA REACLIB, which are from \citet{PhysRevC.85.065809}.
The red and blue lines represent the reaction rates reported in
\citet{2021PhRvC.103a5805A} and \citet{2023EPJA...59...11W}.
Bottom: $^{17}$O($\alpha$, $\gamma$)$^{21}$Ne and $^{17}$O($\alpha$, n)$^{20}$Ne
reaction rates from different references. The gray dashed lines show the default rates
in JINA REACLIB, which are from CF88 and NACRE, respectively.
The red and green lines represent the reaction rates reported in
\citet{2013PhRvC..87d5805B} and \citet{2024PhRvL.132r2701H}.
To show the difference, all the rates shown here are normalized by
the default one in JINA REACLIB.
\label{fig:rate1}}
\end{figure*}

\noindent {\bf(1) From C to Zn:}
With the new $^{22}$Ne+$\alpha$ reaction rates, we observe the 
increase in the yields of several neutron-rich isotopes of Ne, Mg, Si, 
S, Ar, and Ca, particularly those in $^{25}$Mg, $^{29, 30}$Si, $^{36}$S, 
$^{40}$Ar, and $^{46}$Ca.
In Figure \ref{fig:rate1} (b), the significant decrease in the 
$^{22}$Ne($\alpha$, $\gamma$)$^{26}$Mg reaction rate at 
$T \simeq 1.5 - 2.0$ GK (during Ne burning) reduces the yield
of $^{26}$Mg. As a result, a greater amount of $^{22}$Ne is
converted to $^{25}$Mg, leading to a significant increase in
neutron production during Ne burning.
Consequently, the yields of $^{21, 22}$Ne increase due to neutron    
capture on $^{20}$Ne. Similarly, some $^{29, 30}$Si and most of the
rare isotopes $^{36}$S, $^{40}$Ar and $^{46}$Ca are also produced
in Ne burning shell \citep{1995ApJS..101..181W}.
On the contrary, the new $^{17}$O+$\alpha$ reaction rates do not increase
neutrons in the Ne shell, thus leaving the yields of those isotopes
unchanged. The iron peak isotopes remain unaffected by both new 
$^{17}$O+$\alpha$ and $^{22}$Ne+$\alpha$ reaction rates,
as they are primarily produced by the NSE process
at $M_r \simeq 1.8 - 2.2$ M$_\odot$, which is affected by
$Y_e$. However, $Y_e$ is only altered by the weak interaction. 

\noindent {\bf(2) From Ga to Zr:}
Isotopes in this range are most significantly changed by the new
reaction rates. The effect of the new $^{17}$O+$\alpha$ rates significantly differs from
that of the new $^{22}$Ne+$\alpha$ reaction rates.
For the same element, the increases due to the new $^{17}$O+$\alpha$
reaction rates in the isotopic yields are similar.
However, for the new $^{22}$Ne+$\alpha$ reaction rates,
the yields of some isotopes with fewer neutrons are reduced.
With enriched neutrons, the isotopic yields increase quickly.
As $M(\rm ZAMS)$ increases, the enhancement due to the
new $^{17}$O+$\alpha$ reaction rates also increases,
while the enhancement due to the new $^{17}$O+$\alpha$ reaction
rates is not obviously affected by $M(\rm ZAMS)$.

\noindent {\bf(3) From Mo to Bi}
Isotopes in this range are not significantly altered by the new reaction
rates. The new $^{22}$Ne+$\alpha$ reaction rates increase the
yields of only a few isotopes for $M(\rm ZAMS) = 15$ M$_\odot$ models.      
The number of such sotopes increases only slightly for more massive models.

\begin{figure}[htbp]
\centering
\begin{minipage}[c]{0.48\textwidth}
%\centerline{(a) Default Rates}
\includegraphics [width=85mm]{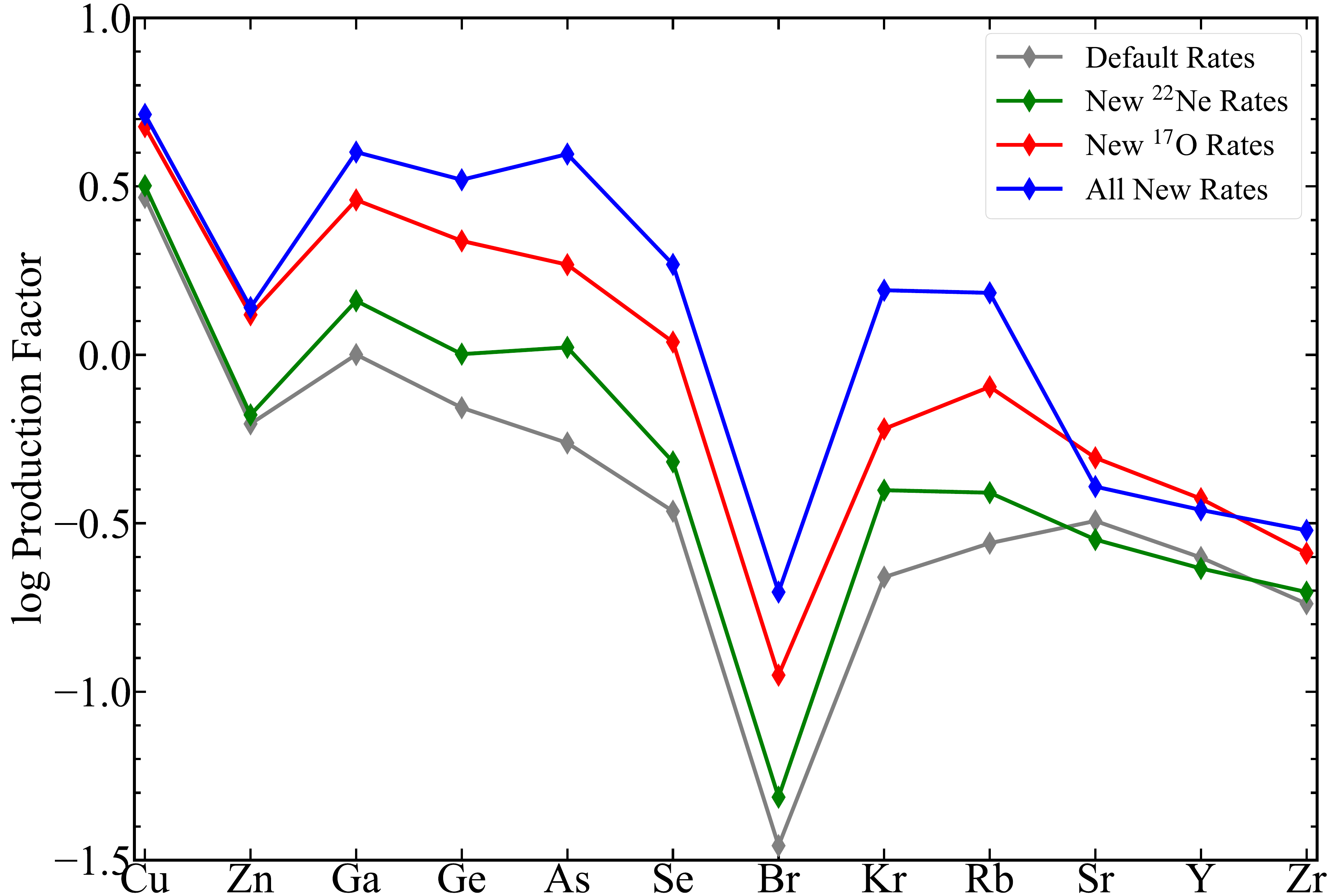}
\end{minipage}%
\caption{The production factors of the elements from Cu to Zr.
Each element is integrated from 15 to 30 M$_\odot$ with the Salpeter
IMF with $\gamma=-2.35$.
\label{fig:yields_element}}
\end{figure}

In principle, the abundance of elements in the solar system arises from
the cumulative contributions of numerous generations of stars with varying
metallicities. Typically, the stars in the range $0.1 Z_\odot < Z < Z_\odot$
contribute more than 90\% to the solar abundance \citep{2003ApJ...592..404L}.
It is general to use the production factors (PFs) to see which element
is contributed significantly by a generation of stars.
The PF of element $i$ is defined as:

\begin{equation} \label{equ:pf}
P_i \equiv \frac{Y^*_i}{X_{i\, \odot} \cdot \sum_k Y^*_i},
\end{equation}
where $X_{i\, \odot}$ denotes the solar mass fraction of element $i$.
$\sum_k Y^*_i$ runs over all elements and
$Y^*_i$ represents the yield of element $i$ averaged by the initial mass
function (IMF) from \citet{1955ApJ...121..161S} with $\gamma=-2.35$. 

Figure \ref{fig:yields_element} shows the production factors of the elements
from Cu to Zr. 
Using the default and new $^{22}$Ne+$\alpha$ reaction rates,
most elements are underproduced (green lines). However, when the new $^{17}$O+$\alpha$
reaction rates are included, the PFs from Zn to Rb
increase by more than a factor of 0.5 dex.
The contribution of stars with 0.1 $Z_\odot$ to the solar
abundance should be limited, as approximately 50\% of the solar
abundance is produced from stars with $0.5\, Z_\odot < Z < Z_\odot$.
Nevertheless, accounting for both new reaction rates leads to
overproduction of Ga, Ge, As, and Se.
Therefore, it is worth calculating the models with 0.5 $Z_\odot$ 
and $Z_\odot$ and verify whether the predictions concerning these
elements align with observational data.
Given the considerable uncertainties involved,
it is essential to enhance the measurement accuracy of the
($\alpha$,n) reaction rate, especially for the $^{17}$O+$\alpha$ reactions.

\begin{figure*}[htbp]
\centering
\begin{minipage}[c]{0.45\textwidth}
%\centerline{(a) Default Rates}
\includegraphics [width=80mm]{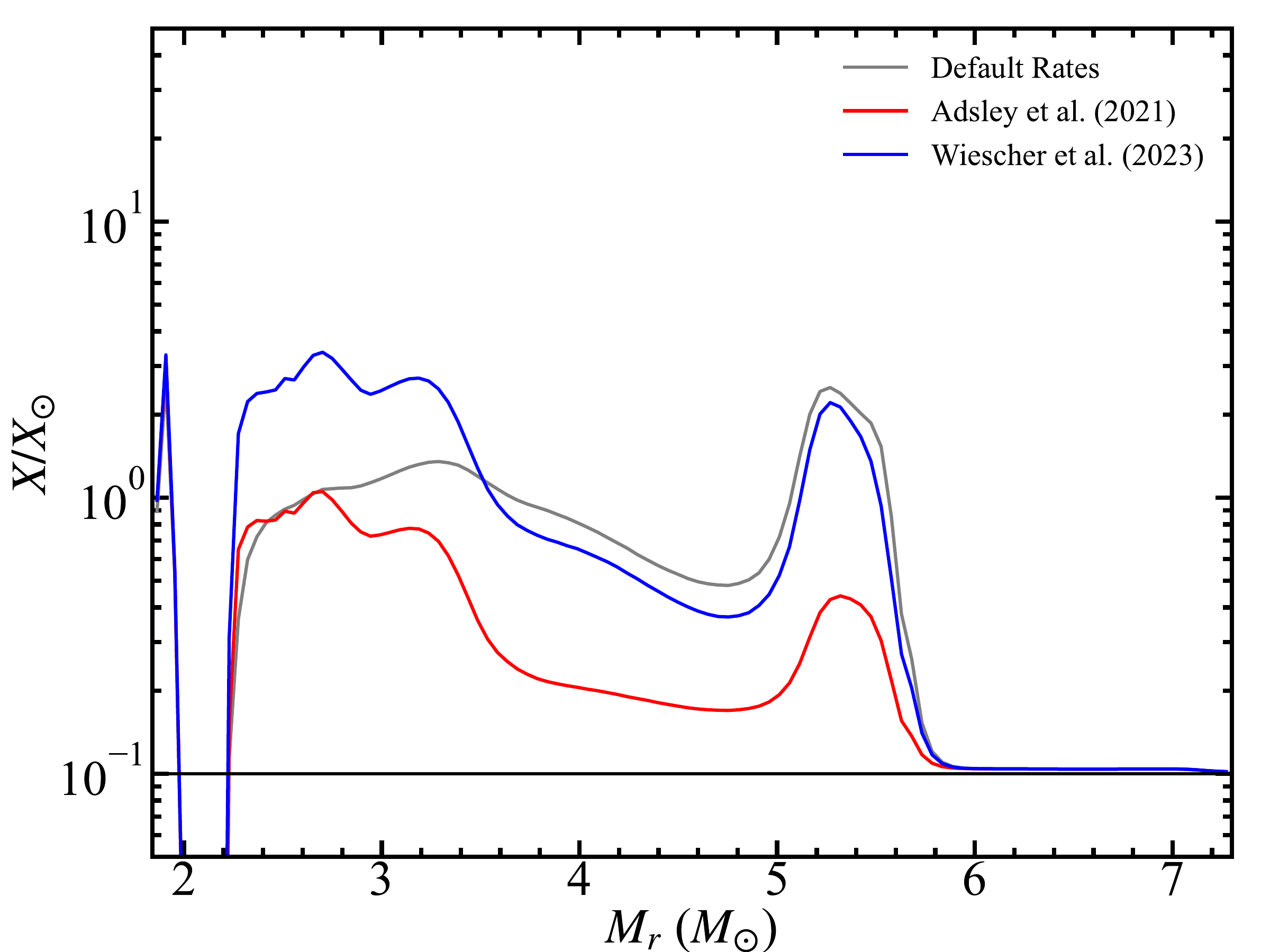}
\end{minipage}%
\begin{minipage}[c]{0.45\textwidth}
%\centerline{(a) Default Rates}
\includegraphics [width=80mm]{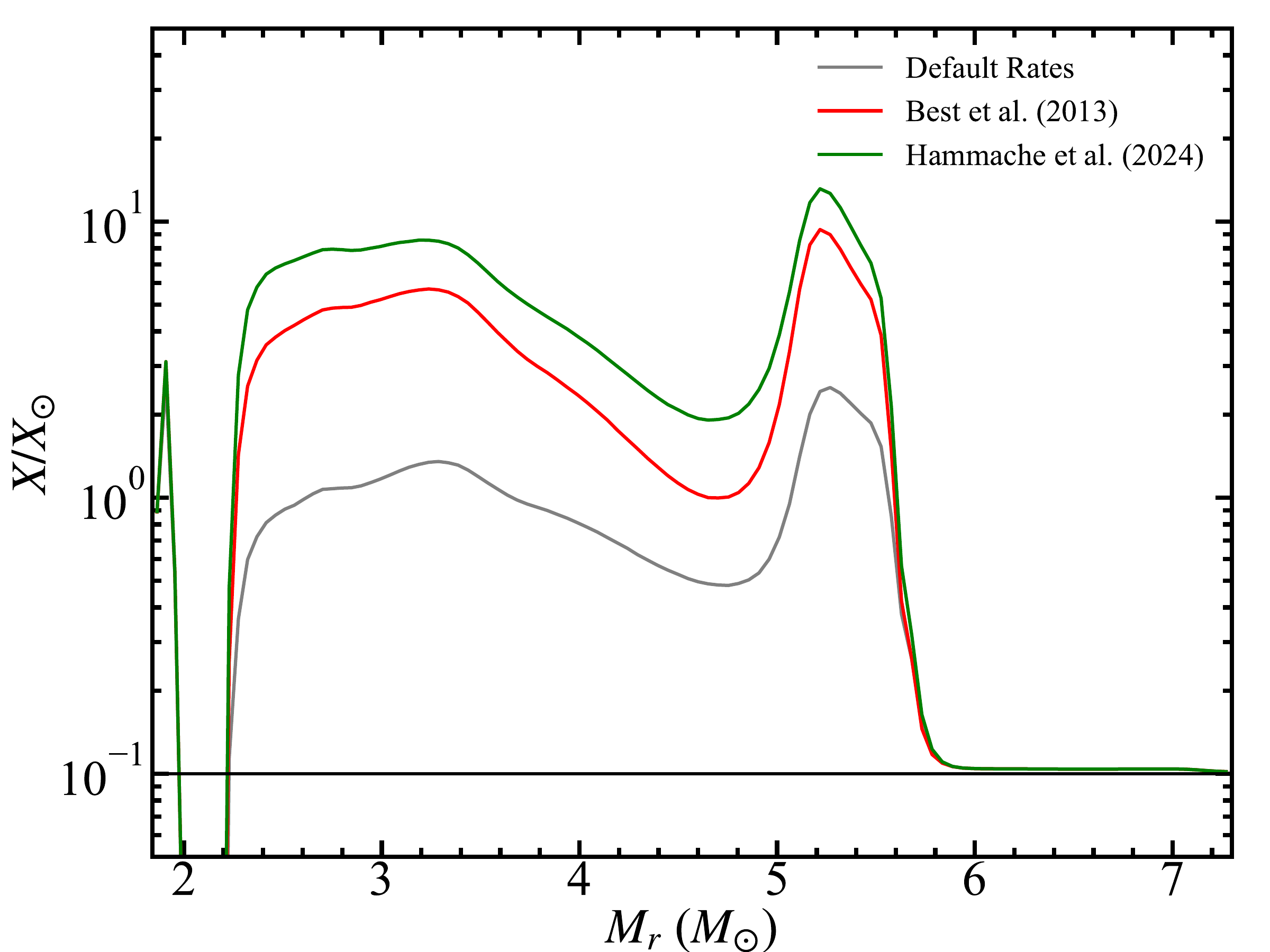}
\end{minipage}%
\caption{Same as Figure \ref{fig:pre2}, but only for the abundances of
``Ga - Zr'' elements.
The effect of $^{22}$Ne+$\alpha$ reaction rates in different
references are compared in the left panel,
while the effect of $^{17}$O+$\alpha$ reaction rates in different
references are compared in the right panel.
\label{fig:diss}}
\end{figure*}

\section{Discussion} \label{sec:diss}

The primary purpose of this work is to evaluate
the effects of the new $^{22}$Ne+$\alpha$ from
\citet{2023EPJA...59...11W} (hereafter W23) and $^{17}$O+$\alpha$
reaction rates from \citet{2013PhRvC..87d5805B} (hereafter B13).
In fact, some other recent reaction rates have been reported by 
\citet{2021PhRvC.103a5805A} (hereafter A21) and
\citet{2024PhRvL.132r2701H} (hereafter H24),
which are not adopted in our evaluations. In Section \ref{sec:diff_rate}, 
we will briefly discuss the effects of differences in these reaction
rates on our $M(\rm ZAMS)$ = 25 M$_\odot$ model.
Besides, the size of the nuclear network used in MESA is limited to $\sim$300
isotopes, which makes it challenging to cover all the s-process isotopes.
Therefore, the evolution and trajectories are based on the MESA calculation
but the detailed nucleosynthesis is based on WinNet.
In this section, we will discuss some uncertainties related to our calculation.

\subsection{Comparisons with Results from Other Reaction Rates}
\label{sec:diff_rate}

In Figure \ref{fig:rate1} (a), we compare the $^{22}$Ne+$\alpha$ and
$^{17}$O+$\alpha$ reaction rates used in the JINA REACLIB with
those reported in different references. As we have mentioned in
\ref{sec:rates}, the ($\alpha$, n)/($\alpha$, $\gamma$) ratio 
of W23 increases quickly above 1.5 GK, because the 
$^{22}$Ne($\alpha$, $\gamma$)$^{26}$Mg reaction rate decreases significantly.
Compared with the default rates, the rates in W23 enhances the contribution
of Ne burning in the ws-process (see Figure \ref{fig:burn_enhance}).
Related enhancement is also observed in the abundance of ``Ga - Zr'' 
elements in the region of $M_r =$ 2.2 - 3.5 M$_\odot$ in Figure \ref{fig:diss}. 

Because A21 only provides the reaction rates below 1.25 GK, 
in this trial, we switch to the rate from \citet{PhysRevC.85.065809} at 
$T > 1.25$ GK. In Figure \ref{fig:rate1} (b),
A21 suggested a lower $^{22}$Ne($\alpha$, n)$^{25}$Mg reaction rate
at $T<0.7$ GK. As a result, the production of the ``Ga - Zr'' elements
is reduced in both core and shell He burning, as shown in Figure \ref{fig:diss}. 

In Figure \ref{fig:rate1} (c) and (d), H24 suggests a similar
$^{17}$O($\alpha$, $\gamma$)$^{21}$Ne reaction rate with the rate in B13
and a higher $^{17}$Ne($\alpha$, n)$^{20}$Ne reaction rate at $T<$ 0.7 GK.
At $T>$ 1.0 GK, we also switch to the rate in B13.
As a result, the production of the ``Ga - Zr'' elements during core He burning
is enhanced compared with the model with the B13 rate in Figure \ref{fig:diss}. 

In conclusion, the $^{22}$Ne+$\alpha$ reaction rates reported in different
studies lead to variations in the yields of the ws-process elements by a factor
of about 3 - 5. In contrast, the $^{17}$O+$\alpha$ reaction rates published
in the literatures produce an order of magnitude differences in
the ws-process elemental yields.

\subsection{The Effect of Mixing} \label{sec:mix}

Nucleosynthesis in each zone is calculated separately
because WinNet is a one-zone code. Thus, the effect of convective
mixing is not taken into consideration. The mixing affects our results
mainly in two aspects. As seen in Figure \ref{fig:dist_abun},
$^{22}$Ne and $^{17}$O are exhausted only in burning shells in the CO
core, because the abundance of $\alpha$ particles is quite small in the unburned
shells. While in Figure \ref{fig:25M_isos} in the region of $M_r$ = 2.5 - 5.1 M$_\odot$,
mixing can transport $^{22}$Ne and $^{17}$O from unburned shells to
the burning shells. Consequently, more neutrons should be released in
MESA calculation.
Similarly, the mass fraction of $^{17}$O in Figure \ref{fig:25M_isos} is quite low.

The convective mixing can also affect the locations of
C, Ne, O, and Si burning shell. With mixing, more fresh fuels are
transported from the outer layers to the bottom of the burning shells.
This would make the life of burning shells longer. 
Thus, the bottom of shell O burning 
located at $M_r$ = 1.84 M$_\odot$ in Figure \ref{fig:25M_isos},
would move to $M_r$ = 2.10 M$_\odot$ in Figure \ref{fig:dist_abun}.
Similarly, the bases of the C and Ne shells at $M_r$ = 2.0 M$_\odot$
would shift to $M_r$ = 2.6 and 2.7 M$_\odot$ without mixing in WinNet.

\subsection{The Effect of Explosion} \label{sec:exp}

As mentioned in Section \ref{sec:exp_pre}, we assume that the mass 
cut locates at $M(V/U_{\rm max})$ and the region of $M_r> M(V/U_{\rm max})$
contributes to the chemical enrichment. We also assume that the ws-process
isotopes produced in the explosive nucleosynthesis would be destroyed by
the shock during the explosion. Thus, we don't calculate the explosive
nucleosynthesis. The results from \citet{2009ApJ...702.1068T} show that
the explosive burning would reduce the ws-process isotopes by less than 15\%.
\citet{2003ApJ...592..404L} mentioned that for the isotopes of $^{70}$Zn,
$^{76}$Ge, $^{74, 77, 82}$Se, $^{78}$Kr, $^{87}$Rb, and $^{84}$Sr, more than
50\% of the yields are produced during explosive burning.
These isotopes should be changed significantly by the shock wave. 
Since the exact explosion mechanism of core-collapse supernovae has
not been well understood, the explosion energy and the choice of the
mass cut would also affect the final yields of those isotopes.

\subsection{Other Effects} \label{sec:other}

Since the ws-process takes place mainly during the He, C, and Ne burning phases,
The physical processes that affect these burning phases may also affect the
ws-process yields, such as reaction rates, convection, rotation, and magnetic
fields \citep{Hirschi2023}.
\citet{2009ApJ...702.1068T} have shown a 15\% change in the 3 $\alpha$ and
$^{12}$C($\alpha$, $\gamma$)$^{16}$O reaction rates may change the yields
of the ws-process isotopes by more than a factor of 2.
\citet{Limongi_2018} presented a large number of rotating massive star models
including the ws-process nucleosynthesis. Their models involve 
$M(\rm ZAMS) =$ 13 - 120 M$_\odot$ and metallicity of
-3 $\leq [\mathrm{Fe}/\mathrm{H}] \leq$ 0.
They find the interplay between the He core and the H burning shell, which 
triggered by the rotation-induced instabilities, enhances the products of
CNO (especially for $^{14}$N) and produces more neutrons.
As a result, the ws-process should be more significantly enhanced in rotating
models.

\section{Conclusion} \label{sec:conclusion}

In this work, we investigate the impact of new $^{17}$O+$\alpha$ reaction rates
from \citet{2013PhRvC..87d5805B} and new $^{22}$Ne+$\alpha$ reaction rates
from \citet{2023EPJA...59...11W} in comparison to the default reaction rates
in JINA REACLIB. We calculate nucleosynthesis of approximately 2000 isotopes, 
ranging from neutron and proton to thorium ($Z=90$)
using the one-zone code WinNet and the stellar models are calculated with MESA
for an initial metallicity of $Z=0.1$ $Z_\odot$ and 
$M(\rm ZAMS) =$ 15, 20, 25 and 30 M$_\odot$.
All the models are evolved from ZAMS to the Fe core collapse, where the infall
speed of the Fe core reaches 10$^3$ km s$^{-1}$. 
We assume that the corrections by explosive nucleosynthesis to the yields are
minor and the isotopes lying in the outer layer of the mass cut ($M_r> M(V/U_{\rm max})$)
would contribute to the chemical enrichment of the Galaxy.
The results are summarized as follows.

(1) The new $^{22}$Ne+$\alpha$ reaction rates slightly suppress the
ws-process during He burning,
while the new $^{17}$O+$\alpha$ reaction rates have the opposite effect.
Both of them enhance the ws-process significantly during C burning
and Ne burning. 
Using the reaction recipes listed in Table \ref{tab:recipes},
$X$ (Ga-Zr) increases by a factor of 6.56, 23.77, 31.58 and 113.62, respectively, 
after Ne burning (see Figure \ref{fig:burn_enhance}).

(2) We note that without considering the effect of mixing,
the mass distribution of the ws-process isotopes provided by WinNet
shows a two-bump shape (see Figure \ref{fig:pre2}).
This is because the unburned layers release fewer neutrons than
the burning shells.
This results in the underestimation of the yields of the ws-process isotopes.
If nucleosynthesis of the ws-process is calculated, coupling with
evolution instead of post-processing, the enhancement of the ws-process
should be more significant.

(3) The new $^{17}$O+$\alpha$ reaction rates can increase the yields of
all isotopes from Cu and Zr, with the enhancement being more
pronounced in more massive stars.
Conversely, the new reaction rates for $^{22}$Ne+$\alpha$
significantly enhance only the yields of the most neutron-rich isotopes
(see Figure \ref{fig:yields}).

(4) We average these four initial masses with Salpeter's IMF and show
the production factors (PFs) of the elements from Cu to Zr.
The new $^{17}$O+$\alpha$ reaction rates enhance the PFs more significantly
than the new $^{22}$Ne+$\alpha$ reaction rates, especially for Ga, Ge, As, and Se.
Considering such a significant impact that the reaction rates from
\citet{2013PhRvC..87d5805B} and JINA REACLIB have on the PFs of these elements,
it is crucial to improve the accuracy and reliability of the measurement of the
$^{17}$O+$\alpha$ reaction rates.
Additionally, further investigations are necessary to ascertain which reaction
rate can explain the astronomical observations well.

(5) We compare the results from the $^{22}$Ne+$\alpha$ and
$^{17}$O+$\alpha$ reaction rates used in the JINA REACLIB with
those reported in different references.
We conclude that the $^{22}$Ne+$\alpha$ reaction rates reported in different
studies lead to variations in the yields of the ws-process elements by a factor
of about 3 - 5. In contrast, the $^{17}$O+$\alpha$ reaction rates published
across the literatures produce an order of magnitude differences in the
ws-process elemental yields.
Because both the $^{22}$Ne+$\alpha$ and $^{17}$O+$\alpha$ reaction rates
have important contributions of the ws-process after core He burning, 
we suggest that researchers conducting experimental or
theoretical studies of nuclear reaction rates provide reaction rates
spanning the temperature range of 0.1 - 10 GK.


\bibliography{sample631}{}
\bibliographystyle{aasjournal}



\end{document}